\documentclass[fleqn,usenatbib,useAMS]{mnras}

\usepackage{newtxtext,newtxmath}

\usepackage[T1]{fontenc}
\usepackage{ae,aecompl}

\usepackage{graphicx}	
\usepackage{subcaption}   
\usepackage{amsmath}	
\usepackage{amssymb}	
\usepackage[separate-uncertainty=true,multi-part-units=brackets,print-unity-mantissa=false]{siunitx}      
\usepackage{commath}      
\usepackage{physics}  
\usepackage[noabbrev]{cleveref}     
\usepackage[dvipsnames]{xcolor}

\newcommand*\mean[1]{\langle #1 \rangle}
\newcommand{\HII}{\ion{H}{II}} 
 
\newcommand{\sigmasfr}{\Sigma_\mathrm{SFR}} 
\newcommand{\sigmagas}{\Sigma_\mathrm{gas}} 
\newcommand{\rhosfr}{\rho_\mathrm{SFR}} 
\newcommand{\rhogas}{\rho_\mathrm{gas}}

\newcommand{\bsda}{bar} 
\newcommand{\bsaa}{inner arm} 
\newcommand{\bsca}{outer arm} 
\newcommand{\bsga}{inter-arm} 

\usepackage[normalem]{ulem}
\usepackage[dvipsnames]{xcolor}

\crefformat{equation}{\eqA equation~\eqB #2#1#3)}
\newcommand{\eqA}{}
\newcommand{\eqB}{(}

\crefname{figure}{Fig.}{Figs.}
\Crefname{figure}{Fig.}{Figs.}
\crefname{table}{Table}{Tables}
\Crefname{equation}{Equation}{Equations}
\newcommand{\solar}{\ensuremath{_{\odot}}} 
\DeclareSIUnit\msol{M\solar{}}
\DeclareSIUnit\lsol{L\solar{}}
\DeclareSIUnit\rsol{R\solar{}}
\DeclareSIUnit\zsol{Z\solar{}}
\DeclareSIUnit\erg{erg}
\DeclareSIUnit\jy{Jy}
\DeclareSIUnit\yr{yr}
\DeclareSIUnit\micron{\micro\metre}
\DeclareSIUnit\au{au}
\DeclareSIUnit\pc{pc}
\DeclareSIUnit\sr{sr}
\DeclareSIUnit\str{sr}
\DeclareSIUnit\kms{\km\per\s}
\DeclareSIUnit\Myr{\mega\yr}

\definecolor{arpcolor}{RGB}{0, 153, 76}

\title[Star formation in different galactic environments]{Star cluster formation and feedback in different environments of a Milky Way-like galaxy}

\author[A. A. Ali, et al.]{
Ahmad A. Ali$^{1,2}$\thanks{E-mail: ahmadali@ph1.uni-koeln.de}, 
Clare L. Dobbs$^{1}$,
Thomas J. R. Bending$^{1}$,
Anne S. M. Buckner$^{1}$ and
Alex R. Pettitt$^{3}$
\\
$^{1}$Department of Physics and Astronomy, University of Exeter, Stocker Road, Exeter EX4 4QL, United Kingdom \\
$^{2}$I. Physikalisches Institut, Universit\"{a}t zu K\"{o}ln, Z\"{u}lpicher Str. 77, 50937 K\"{o}ln, Germany \\
$^{3}$Department of Physics and Astronomy, California State University, Sacramento, 6000 J Street, Sacramento, CA 95819-6041, USA
}

\date{Accepted 2023 June 20. Received 2023 June 19; in original form 2023 March 09}

\pubyear{2023}

\begin{document}
\label{firstpage}
\pagerange{\pageref{firstpage}--\pageref{lastpage}}
\maketitle

\begin{abstract}
It remains unclear how galactic environment affects star formation and stellar cluster properties. This is difficult to address in Milky Way-mass galaxy simulations because of limited resolution and less accurate feedback compared to cloud-scale models. 
We carry out zoom-in simulations to re-simulate 100--\SI{300}{\pc} regions of a Milky Way-like galaxy using smoothed particle hydrodynamics, including finer resolution (\SI{0.4}{\msol} per particle), cluster-sink particles, ray-traced photoionization from O stars, H$_2$/CO chemistry, and ISM heating/cooling. We select $\sim$\SI{e6}{\msol} cloud complexes from a galactic bar, inner spiral arm, outer arm, and inter-arm region (in order of galactocentric radius), retaining the original galactic potentials. 
The surface densities of star formation rate and neutral gas follow  $\sigmasfr \propto \sigmagas^{1.3}$, with the bar lying higher up the relation than the other regions. However, the inter-arm region forms stars 2--3x less efficiently than the arm models at the same $\sigmagas$. The bar produces the most massive cluster, the inner arm the second, and the inter-arm the third. Almost all clusters in the bar and inner arm are small (radii $< \SI{5}{pc}$), while 30-50 per cent of clusters in the outer arm and inter-arm have larger radii more like associations. Bar and inner arm clusters rotate at least twice as fast, on average, than clusters in the outer arm and inter-arm regions. The degree of spatial clustering also decreases from bar to inter-arm. Our results indicate that young massive clusters, potentially progenitors of globular clusters, may preferentially form near the bar/inner arm compared to outer arm/inter-arm regions.
 
\end{abstract}

\begin{keywords}
stars:formation -- HII regions -- ISM: clouds -- galaxies: star formation -- galaxies: star clusters: general
\end{keywords}




\section{Introduction}
Star formation takes place in giant molecular clouds (GMCs) with most stars forming in clusters or associations \citep{lada2003}. How these clusters/associations form is still an open problem, as is the cause of differences in their properties. Young massive clusters (YMCs; masses $>$\SI{e4}{\msol}, radii $\sim$\SI{1}{pc}) are of particular interest as they may be the progenitors of globular clusters \citep{portegies-zwart2010,longmore2014}. In addition to the cluster properties, the gas itself is influenced by the newly formed stars. Once massive stars ($>$\SI{8}{\msol}) form, they become a source of feedback by releasing energy and momentum into the interstellar medium (ISM), changing the gas dynamics while stars are still forming. This can affect star formation rates by dispersing gas reservoirs \citep{walch2012} or by compressing them to form new stars \citep{elmegreen1977,whitworth1994}. 

Star formation, feedback, and cluster properties may depend on their birth environment. The observed relation between gas surface density and star formation rate (SFR) surface density depends on galactocentric radius, with SFRs being higher at smaller radii \citep{bigiel2008}. 
Numerical simulations have extensively shown that feedback depends on initial cloud conditions such as mass \citep{dale2014,howard2017a,ali2019}, surface density \citep{kim2018a,fukushima2022}, metallicity \citep{fukushima2020,ali2021}, structure \citep{walch2013,zamora-aviles2019}, gravitational boundedness \citep{howard2016}, and turbulence \citep{geen2018,guszejnov2022}. 
The most important feedback mechanism on cloud scales appears to be photoionization, which heats gas from $\sim$\SI{e2}{K} to \SI{e4}{K}, creating a pressure gradient between ionized gas and the neutral ISM. While there are still many uncertainties, photoionization may dominate over other pre-supernova (SN) mechanisms such as stellar winds \citep{geen2021,ali2022} and radiation pressure \citep{kim2018a,ali2021}. These mechanisms set the structure into which SNe explode, potentially creating low-density channels through which energy can escape \citep{lucas2020,bending2022}. 

Results from cloud-scale studies need to be placed in the larger galactic context. GMC evolution is influenced by galaxy-scale potentials, shear, and cloud-cloud tidal forces \citep{dobbs2013,jeffreson2020}. 
Observations in NGC 300 show that feedback-related pressure terms exhibit a slight dependence on galactocentric radius, indicating that feedback becomes more powerful at larger radii \citep{mcleod2021}. \HII{} regions at small galactic radii may be confined by higher ambient pressures \citep{barnes2020,della-bruna2022a}, meaning \HII{} regions in the disc may be able to expand to greater sizes compared to those nearer the centre. 

However, it is less clear how specific galactic structures such as bars and spiral arms affect the star formation and feedback processes. Individual clouds in the Central Molecular Zone of the Milky Way (the innermost 500\,pc) are observed to be significantly less star-forming than expected given their high densities \citep{longmore2013,kauffmann2017}. This may be due to galaxy-scale potentials causing strong shear \citep{kruijssen2019}.  Yet galactic centres do contain young massive clusters, including, in our own Galaxy, the Arches and Quintuplet clusters \citep{longmore2014}. The Galactic bar and its intersections with spiral arms might also host YMCs \citep{davies2012,ramirez-alegria2014}, including W43 \citep{nguyen-luong2011,carlhoff2013}. While disc clouds may have lower masses or mean densities than clouds in the centre, such regions still manage to form YMCs such as NGC 3603, potentially through cloud-cloud collisions \citep{fukui2014,liow2020}.

Numerical simulations which explore these processes can broadly be split into two types: galaxy-scale simulations which model the evolution of a whole Milky Way-mass galaxy over 100's of Myr and follow the interaction between GMCs and large-scale structures such as spiral arms (see the review by \citealt{naab2017}; also, \citealt{agertz2013,dobbs2013a,smith2020,jeffreson2020,pettitt2020,keller2022a}). The second type follows the cloud-scale evolution over 3--10\,Myr and follows star formation and feedback without external influences. In these models, it is computationally feasible to include high-resolution pre-SN feedback methods, but the initial conditions are idealised, usually in the form of turbulent spherical clouds (see the review by \citealt{dale2015}; also, \citealt{walch2013,dale2014,grudic2018,kim2018a,ali2018,geen2018}). 

In this paper, and previous papers \citep[including][]{bending2020,ali2022,dobbs2022,herrington2023}, we attempt to bridge this gap by extracting GMC complexes from galaxy simulations (thus starting with more realistic density and velocity distributions; \citealt{rey-raposo2017}), and using feedback methods such as ray-tracing which are usually limited to smaller scales. These zoom-in simulations retain the galactic potentials and include multiple clouds, thus including tidal forces and shear which isolated cloud models would neglect. By extracting GMC complexes (of mass \SI{e6}{\msol} and size 100--300\,pc) from different parts of a galaxy, we explore the impact of galactic environment on star/cluster formation and feedback. In \cref{sec:numericalmethods}, we describe the zoom-in method and the implementation of star formation and feedback. We present our results in \cref{sec:results}, discussion in \cref{sec:discussion}, and conclusions in \cref{sec:conclusions}.

%
%
\section{Numerical methods}
\label{sec:numericalmethods}

We use the smoothed particle hydrodynamics code \textsc{sphNG}, which originated with \citet{benz1990} and \citet{benz1990a}, with substantial modifications by \citet{bate1995} and \citet{price2007} such as the inclusion of sink particles and magnetic fields (although we do not include the latter here). Our models include self-gravity, ISM heating/cooling, H$_2$ and CO chemistry \citep{glover2007,dobbs2008}, and the galactic potentials described in \cref{sec:initialconditions}. We use the methods introduced by \citet{bending2020} for cluster-sink particles and photoionizing radiation via ray-tracing. We also include supernovae as described by \citet{bending2022}. We summarise the sink method in \cref{sec:clustersinks} and the feedback processes in \cref{sec:feedback}.

\subsection{Initial conditions}
\label{sec:initialconditions}

\begin{figure*}
\begin{subfigure}{0.28\textwidth}
  \centering
  \includegraphics[width=1.0\linewidth]{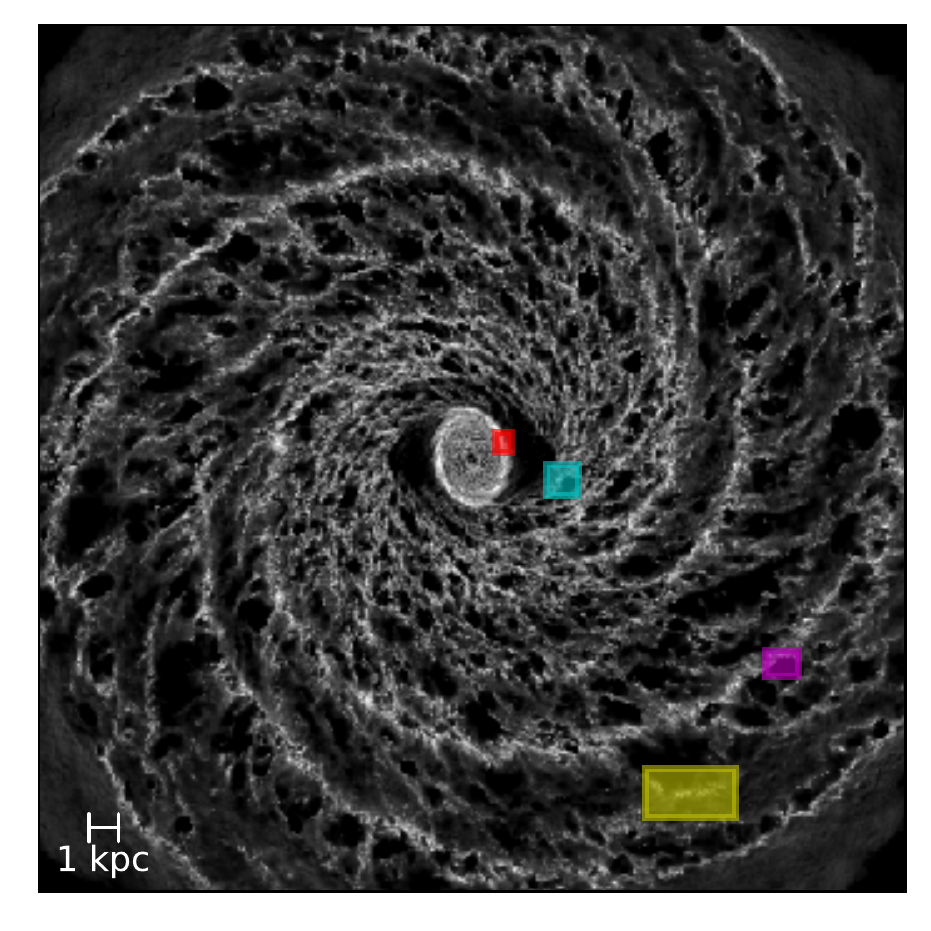}  
\end{subfigure}
\begin{subfigure}{0.70\textwidth}
  \centering
  \includegraphics[width=1.0\linewidth]{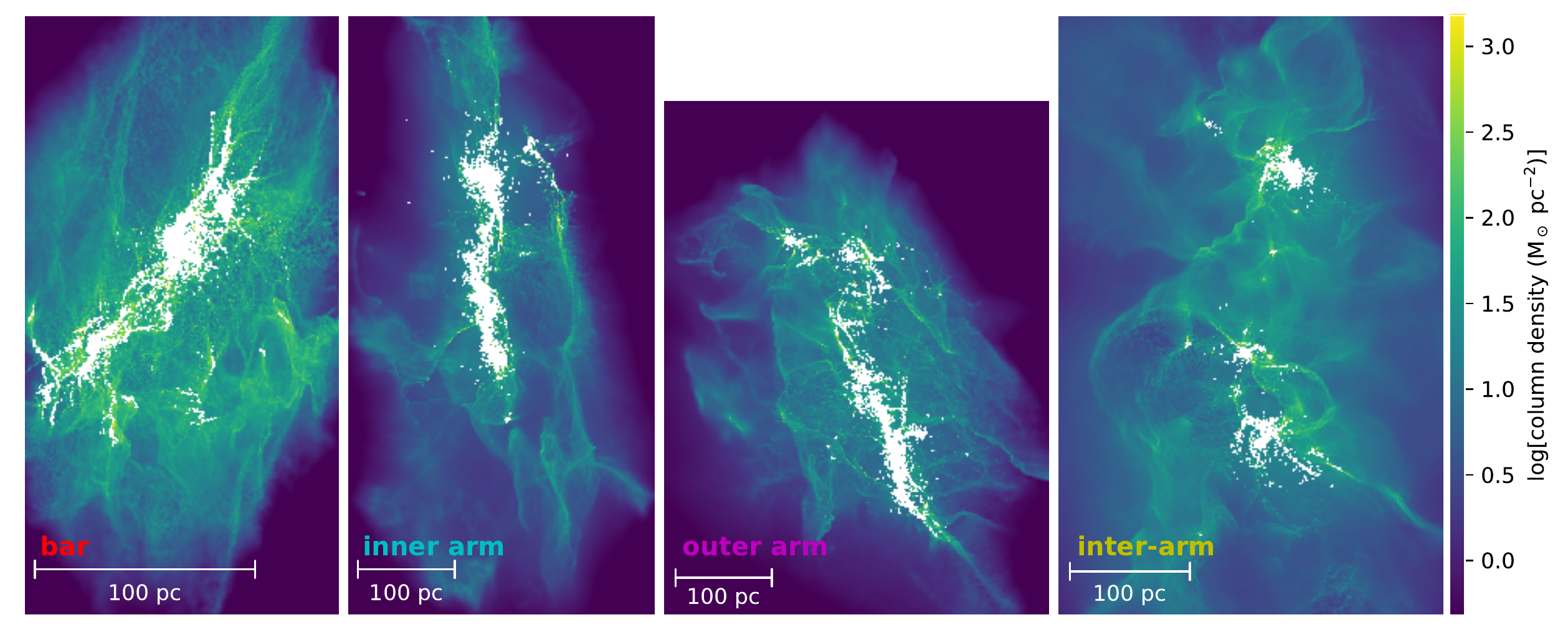}  
\end{subfigure}
\caption{Top-down view of original galaxy model by \citet{pettitt2020} (far left). Regions are taken from the bar, inner spiral arm, outer arm, and inter-arm - these are the zoom-ins which are re-simulated with main sequence feedback in this paper. The galaxy shows the location of initial conditions (the first region enhancement; see \cref{sec:initialconditions}). Figures show column density with sinks in white. 
Snapshots are shown at 2.0\,Myr (\bsda{}),
4.1\,Myr (\bsaa{}),
3.5\,Myr (\bsca{}), and
4.0\,Myr (\bsga{}) after the onset of feedback and have been rotated for this figure.}
\label{fig:initialinset}
\end{figure*}

\begin{table*}
	\centering
	\caption{Initial conditions of zoom-in regions, including galactocentric distance $R_\mathrm{gal}$, total gas mass $M$, number of particles $N_\mathrm{part}$, mean mass per particle $m_\mathrm{part}$, mean density $\mean{\rho}$ in units of the hydrogen mass $m_\mathrm{H}$, and size along each axis.}
	\label{tab:models}
	\begin{tabular}{lcccccc} 
		\hline
		region & 
		$R_\mathrm{gal}$ (kpc) & 
		$M$ (\SI{e6}{\msol}) & 
		$N_\mathrm{part}$  & 
		$m_\mathrm{part}$ (\si{\msol}) & 
            $\mean{\rho}/m_\mathrm{H}$ (\si{\per\cm\cubed}) &
            $X \times Y \times Z$ (pc) \\
		\hline
\bsda{}	 &	0.76 & 1.8 & 4,134,592 & 0.44 & 476 & $122 \times 117 \times 149$\\
\bsaa{}	&	2.1	 & 1.6 & 3,744,488 & 0.43 & 140 & $331 \times 235 \times 291$\\
\bsca{}	&	8.4	 & 1.6 & 3,923,532 & 0.41 & 96.8 & $248 \times 271 \times 327$\\
\bsga{}	&	9.2	 & 2.1 & 4,861,061 & 0.43 & 47.6 & $468 \times 310 \times 348$\\

		\hline
	\end{tabular}
\end{table*}

We set up our initial conditions by extracting a region from a galaxy evolution model, enhancing the resolution, then re-running the region with the higher accuracy methods for sinks and feedback (i.e. a zoom-in simulation). 
The galaxy is based on the Milky Way and includes analytic potentials for a bar and four spiral arms. We use a modified version of the BrSp4 model from \citet{pettitt2020} taken at \SI{340}{\mega\yr} post-initialisation. This simulation was carried out using the SPH code \textsc{gasoline2} \citep{wadsley2004,wadsley2017}. It included stellar feedback in the form of winds from evolved low-mass stars ($<$\SI{8}{\msol}) and supernovae (type II and type Ia; \citealt{stinson2006,pettitt2017}). Main sequence feedback from high-mass stars was not included. We retain the bar and arm potentials in our zoom-in simulations using \textsc{sphNG}. The bar \citep{wada2001} has a scale length of $\sqrt{2}$\,kpc and a pattern speed of \SI{60}{\kms\per\kilo\pc}. The spiral arms \citep{cox2002} have a pitch angle of $15^\circ$ and pattern speed \SI{20}{\kms\per\kilo\pc}. There is also an axisymmetric potential for the combined disc + bulge + halo \citep{pettitt2014}. This model differs to that used in \citet{pettitt2020} in that it includes a greater gas-mass resolution of \SI{600}{\msol} per SPH particle and a gravitational softening scale of 5\,pc (compared to \SI{1500}{\msol} and $50\,$pc presented in the aforementioned paper). The global dynamics and structure are essentially identical to that of the model studied in \citet{pettitt2020} and are not discussed here.

We select gas particles in a region of the galaxy and increase the resolution using the particle-splitting method of \citet{bending2020}. In the following, we use \textsc{sphNG} and retain the original galactic potentials, but we do not take any of the original star particles. For each model, we do the resolution enhancement in two stages. This reduces the effect of grid artefacts in the final particle setup. The first enhancement takes the region (size of the order of $\sim$1\,kpc) and increases the resolution from the base resolution of the galaxy simulation to $\sim$\SI{13}{\msol} per particle. The galactic position of each region at this stage of the process is shown in the left-most panel of \cref{fig:initialinset} overlaid on the original galaxy. We evolve this region (without feedback) for approximately \SI{0.5}{\mega\yr} to allow the particles to settle. Next, we select a sub-region (size 100--300\,pc) and enhance to $\sim$\SI{0.43}{\msol} per particle. This is used as the initial condition to evolve with cluster-sinks and stellar feedback.

The parameters of each initial condition are listed in \cref{tab:models}, sorted by galactocentric distance. One region explored is in the bar, two are in spiral arms (one inner arm and one outer arm), and finally one is in an inter-arm region. The initial mean density is largest for the bar and decreases with galactocentric distance, with the bar being 10 times denser than the outermost region (inter-arm model). The panels to the right of the galaxy in \cref{fig:initialinset} show the models 2--\SI{4}{\mega\yr} after the onset of ionization.

\subsection{Cluster-sink particles}
\label{sec:clustersinks}

The zoom-in simulations form cluster-sink particles which represent (sub-)clusters of stars. Sink formation is based on the criteria  of \citet{bate1995}. Gas particles above a density threshold of \SI{1.2e4}{\per\cm\cubed} are tested to see if the neighbourhood of $\sim 50$ particles is collapsing and converging. If so, the particle and its neighbours are converted to a sink particle. Sink formation is forced for densities above \SI{1.2e6}{\per\cm\cubed}. The sink accretion radius is \SI{0.1}{pc} and the sink merger radius is \SI{0.03}{pc}. 

The method for converting sink masses to a stellar population is based on a method by \citet{geen2018} and is described in \citet{bending2020} with revisions in \citet{herrington2023}. Before the simulation, we create a list of massive stars. This is done by sampling from a \citet{kroupa2001} initial mass function (IMF) up to a total mass of \SI{3e6}{\msol} and grouping stars into mass bins, keeping a list of the bin indices for masses above \SI{18}{\msol}. The same list and ordering is used for all models presented here, as well as in \citet{bending2020} and \citet{ali2022}. We keep track of the mass accreted by each sink. When the total mass accreted over all sinks reaches \SI{305}{\msol}, we take the next massive star from the list and assign its properties (e.g. ionizing flux) to a chosen sink -- this is the sink with the most mass not made up of massive stars. This is only done if the sink is massive enough to accept the star; otherwise, we wait until the next time step.
The fraction of the sink mass that is available for star formation is 50 per cent -- the rest is assumed to be a gas reservoir. Stars are in bins according to spectral type with representative masses and ionizing fluxes listed in table 2 of \citet{bending2020}.

\subsection{Stellar feedback}
\label{sec:feedback}
We use a ray-tracing method for calculating photoionization equilibrium along lines of sight (LOS) between gas particles and sinks. This is a similar method to \citet{dale2007b}. A full description is available in \citet{bending2020}. For each gas particle, we calculate the ionizing flux it receives from all ionizing sources, taking into account the reduction along the LOS. All particles with a smoothing length which overlaps with the LOS are included, with quantities interpolated at the position on the LOS (rather than the particle position, as \citeauthor{dale2007b} do).  The photoionization rate is balanced with the recombination rate at the gas particle density. We  use the on-the-spot approximation with case B recombination coefficient $\alpha_\textrm{B}=\SI{2.7e-13}{\cm\cubed\per\s}$ and ionization temperature of \SI{e4}{K}. If a gas particle receives ionizing radiation from multiple sources, the column density contributions from all lines of sight are divided by the number of sources contributing the flux. This is described in \citet{herrington2023}.

To reduce the computational time, we set the maximum LOS distance to \SI{100}{pc} -- this is tested by \citet{bending2020} in a 500\,pc region. In that case, the longer range ionization only had a small effect on top of the limited LOS model, sweeping up shells near the boundaries where massive stars did not form. Here we investigate smaller regions (100-300\,pc, closer to the LOS radius), and the sink particles are more evenly spread out, meaning we do not expect this limit to change our results significantly. We also group ionizing sinks into nodes if they are close together, which reduces the number of ionizing `sources' that a gas particle needs to loop over. For the most ionizing sink, we find the minimum radius at which the average ionization fraction of enclosed gas drops below 90 per cent. Any other sink that sits within half this radius is grouped with that sink. We repeat with the next most ionizing sink that has not already been grouped, until all sinks have been grouped or tested. The summed flux of each sink group propagates from its centre of flux. We find that this reduces the total number of ionizing sources in the latter stages of our simulations by a factor of between 2 and 3. Testing with and without this optimisation shows only a small effect on the total amount of ionized gas (\textless 1 per cent) and a negligible effect on \HII{} region morphology.

We also include supernovae (SNe) using the method by \citet{dobbs2011}, which was updated for cluster-sink particles by \citet{bending2022}. When a star above \SI{18}{\msol} becomes old enough to explode as a SN, we insert energy around the host sink inside a radius which encompasses its 80 nearest particles. This radius is used to calculate the age, temperature, and velocity of a SN bubble in the snowplough phase, assuming each SN contributes \SI{e51}{erg}. This solution provides the energy to be inserted inside the radius as a combination of thermal and kinetic energy. We do not include SNe from less massive stars as their life times are longer than the simulation run times.

We do not include stellar winds in this paper, as their impact on the gas is negligible compared with photoionization \citep{ali2022}. Their impact on cluster properties may be marginally more important compared with gas properties, so this is planned for future models.

%
%

\section{Results}
\label{sec:results}
\begin{figure*}
\begin{subfigure}{0.24\textwidth}
  \centering
  \includegraphics[width=1.0\linewidth]{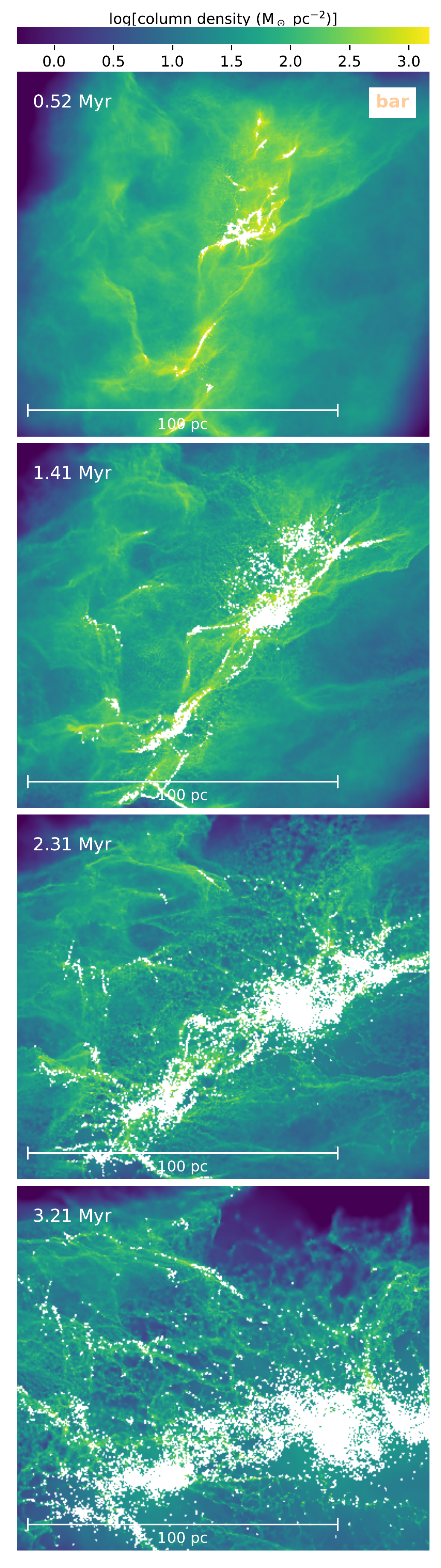}  
\end{subfigure}
\begin{subfigure}{0.24\textwidth}
  \centering
  \includegraphics[width=1.0\linewidth]{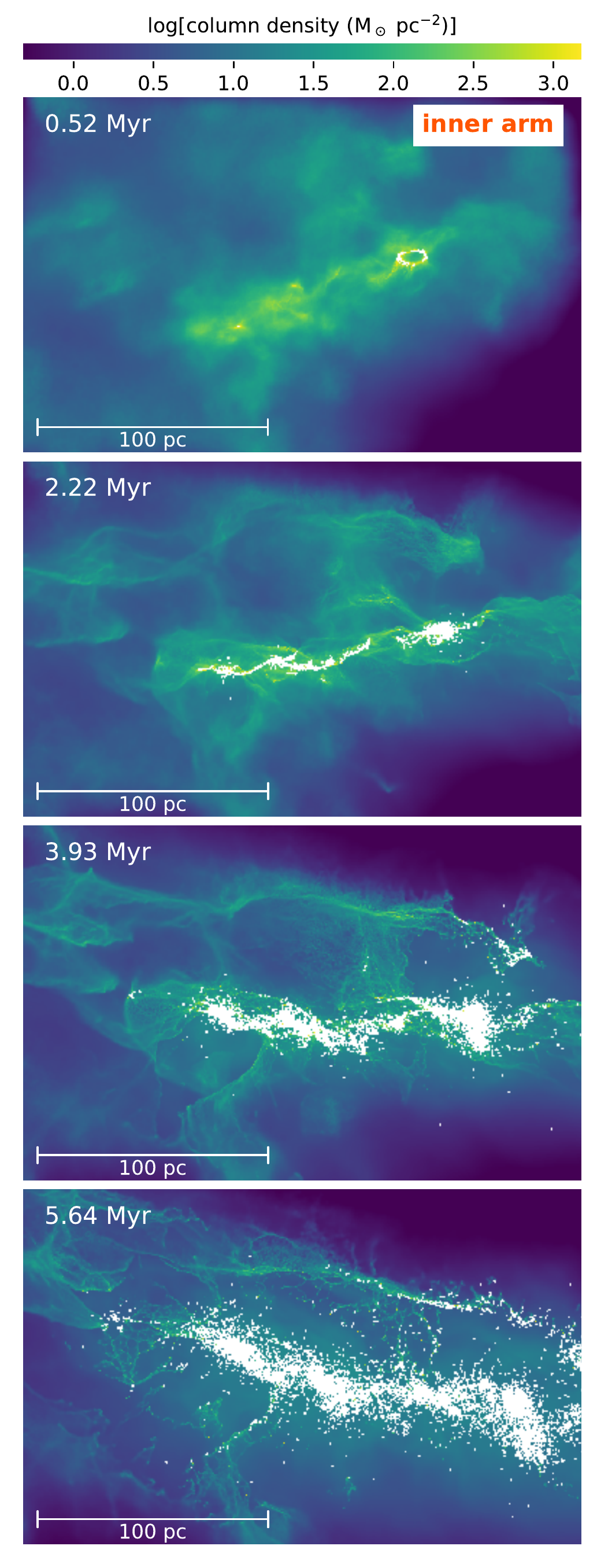}  
\end{subfigure}
\begin{subfigure}{0.24\textwidth}
  \centering
  \includegraphics[width=1.0\linewidth]{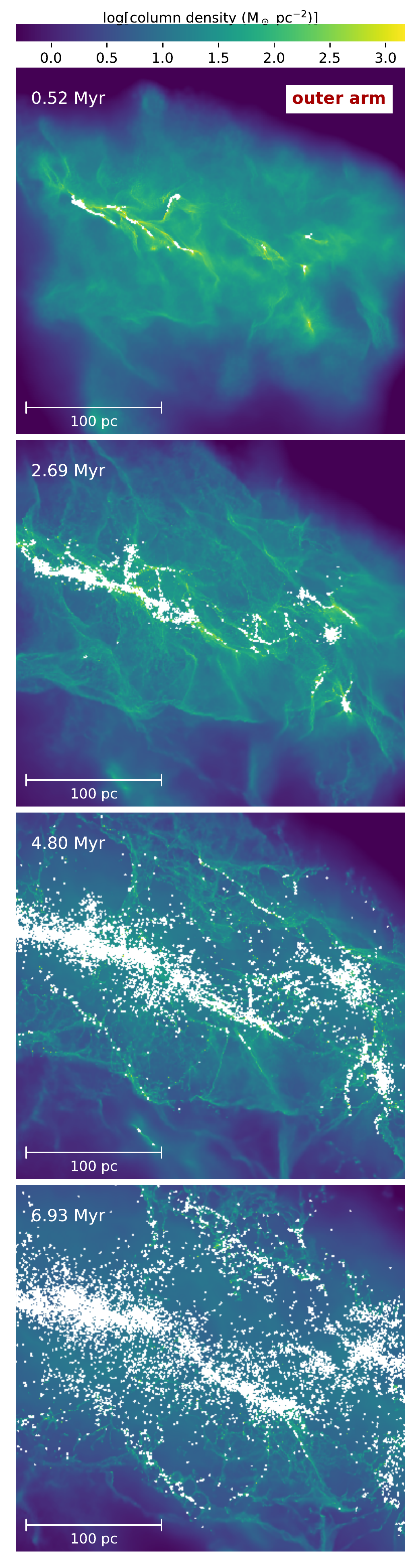}  
\end{subfigure}
\begin{subfigure}{0.24\textwidth}
  \centering
  \includegraphics[width=1.0\linewidth]{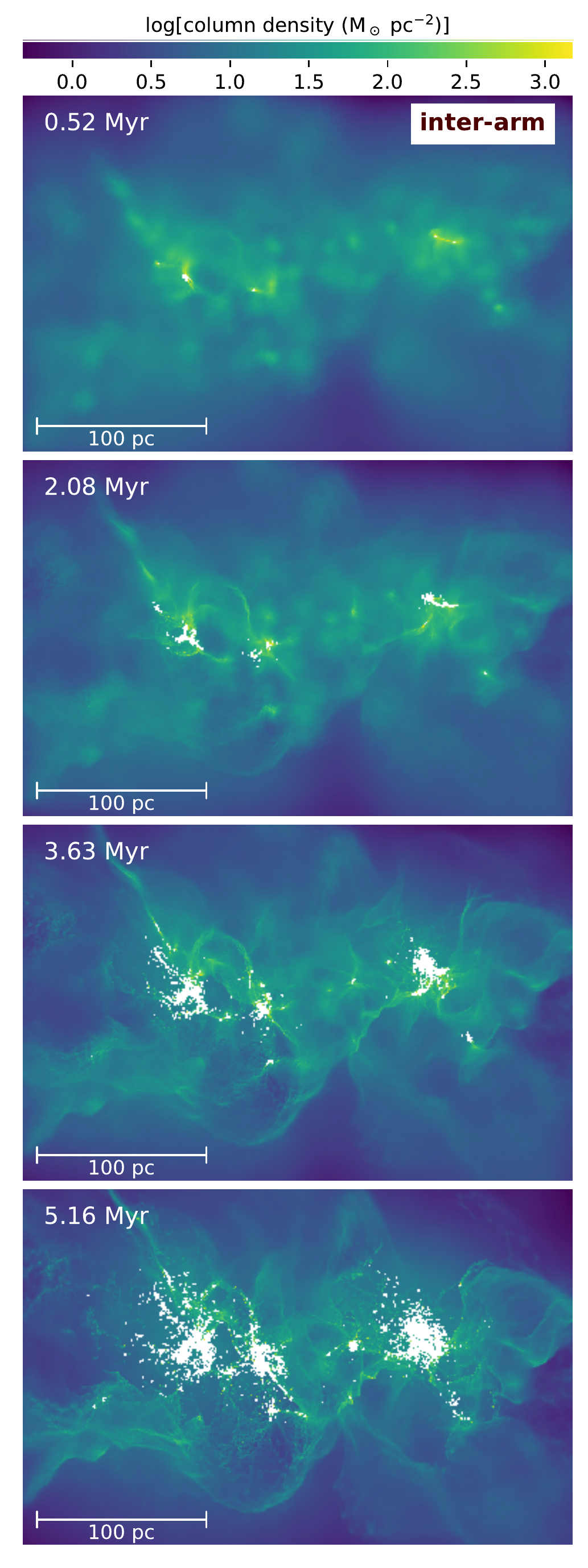}  
\end{subfigure}
\caption{Top-down view of zoom-ins from the bar, inner spiral arm, outer arm, and inter-arm (left to right). Figures show column density with sinks in white. Time evolution is top to bottom -- zero time is the onset of feedback.}
\label{fig:columnevolution}
\end{figure*}

\cref{fig:columnevolution} shows, for each region, the time evolution of column density. Four snapshots are presented between \SI{0.5}{\mega\yr} after the onset of feedback and the end of the simulation run time. The end time is typically limited by the small time steps caused by sink dynamics (in particular the bar model) or heating from SNe. The effect of rotation around the galactic centre can be seen most clearly for the bar region, where the gas structure starts almost vertical and rotates almost 90 degrees over the next 3.2\,Myr. The inner arm also shows some rotation, but this is marginal compared to the bar. The outer arm and inter-arm regions do not show rotation over the time scales modelled here. The bar has a burst of star formation in the first Myr and is dominated by a large, dense cluster which forms from the densest gas. Star formation in the inner arm and outer arm models occurs along the length of the central arm structure, with clusters forming in a chain-like pattern. Star formation in the inter-arm region is more sparsely distributed -- this has three main sites of star formation, with two close together on the left and one $\sim$ 100\,pc away to the right. The inter-arm regions bear the closest resemblance to isolated cloud models.

\subsection{Star formation}
\label{sec:sfr}

\begin{figure}
    \centering
	\includegraphics[width=0.95\columnwidth]{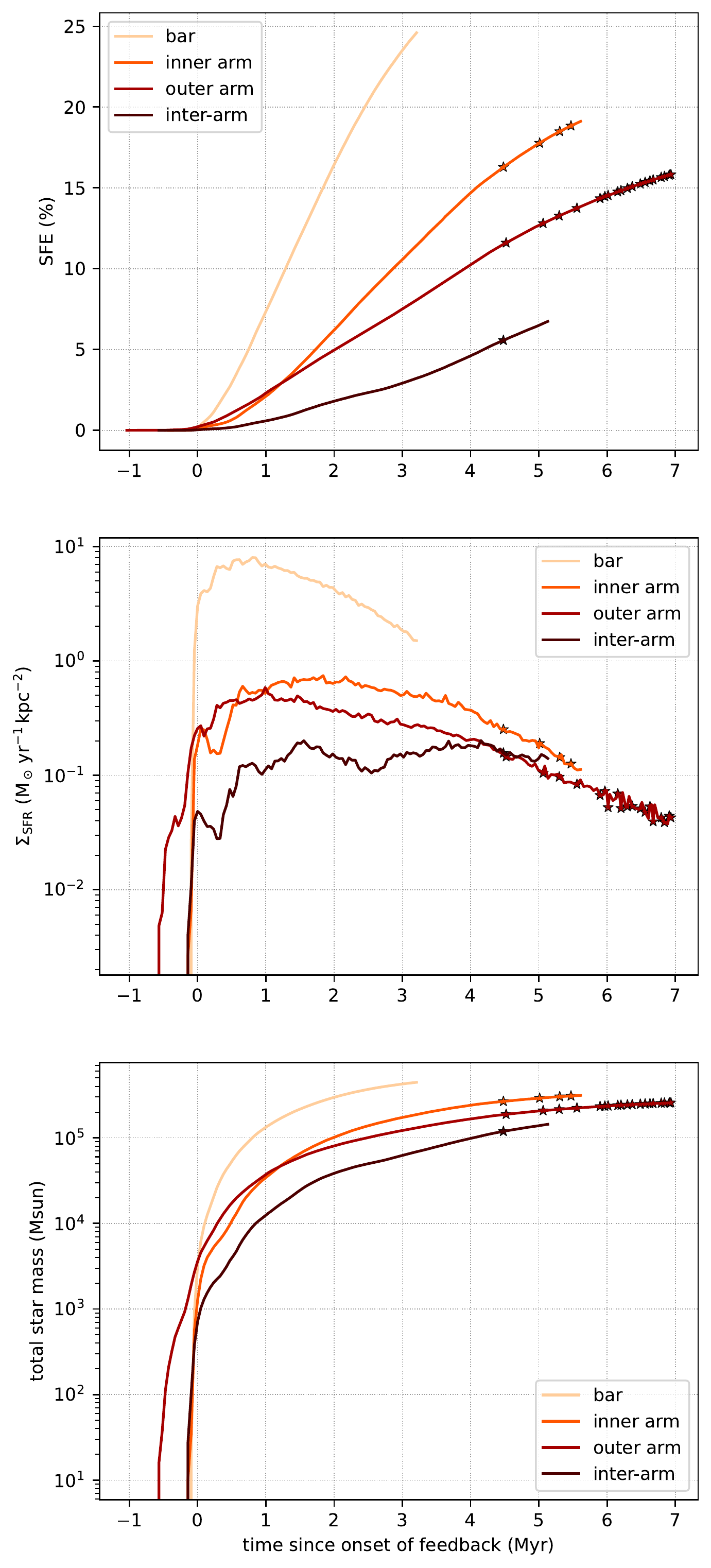}
    \caption{Time-evolution of the star formation efficiency (top), surface density of star formation rate (middle), and total mass in stars (bottom). Star symbols denote supernova events.}
    \label{fig:sfe}
\end{figure}

\cref{fig:sfe} shows the time-evolution of star formation efficiency (SFE), star formation rate surface density ($\sigmasfr{}$), and the cumulative star mass. The SFE is defined in terms of the total sink and gas mass,
\begin{equation}
\mathrm{SFE} = \frac{0.5 M_\mathrm{sinks}}{M_\mathrm{sinks} + M_\mathrm{gas}}~,
\end{equation}
where the stellar mass is half the sink mass according to the cluster-sink prescription described in \cref{sec:clustersinks}. 
$\sigmasfr{}$ is the star formation rate divided by the rectangular area $XY$ which contains 99 per cent of the neutral gas mass, with the origin at the centre of mass, as viewed in the $x$-$y$ (top-down) plane, i.e.
\begin{equation}
    \label{eq:sigmasfr}
    {\sigmasfr} =  \frac{0.5 \Delta M_\mathrm{sinks}}{XY\Delta t}~.
\end{equation}
This is calculated over time intervals $\Delta t = \SI{0.047}{\mega\yr}$ and the dimensions are also re-calculated at every SFR measurement time. $\Delta M_\mathrm{sinks}$ is the change in sink mass over the time interval $\Delta t$. Zero time in the plot is defined to be when ionizing radiation is first emitted. The star symbols show when supernovae explode, the first of which occurs after \SI{4.5}{\mega\yr}.

The final SFE decreases with galactocentric distance, with the \bsda{} model having the highest SFE of 25 per cent by the end of its runtime of 3.2\,Myr. For the models which undergo at least one supernova event, the SFEs just before the first SN (in per cent) are 16 (\bsaa{}), 12 (\bsca{}), and 6 (\bsga{}). This shows that inner regions form proportionately more stars in the same amount of time. For most of the evolution, $\sigmasfr$ is also ordered by galactocentric distance as rates get lower with larger distance. However, this is slightly different in the first Myr, where the two arm regions fluctuate over each other. Similarly, from 4\,Myr, the two outermost regions (the \bsca{} and \bsga{}) overlap before the latter overtakes the former. By 5\,Myr, there is not much variation between the three non-bar regions, with the range in $\sigmasfr$ being within a factor of 2. The peak values of $\sigmasfr$ are reached at 0.8\,Myr (\bsda{}), 1.8\,Myr (\bsaa{}), 1.0\,Myr (\bsca{}), and 4.2\,Myr (\bsga{}); the latter model however shows two peaks of star formation, with the first peak occurring around 1.5\,Myr. The other three models experience a burst of star formation early on, before declining -- star formation is spread more evenly in position across these regions, while the \bsga{} has two distinct sites of star formation separated by $\sim\SI{100}{pc}$. All four models reach total stellar masses above \SI{e5}{\msol}, with the \bsda{} being the first model to achieve this at \SI{0.85}{\mega\yr} and the \bsga{} region being the last at \SI{4.1}{\mega\yr}.

\begin{figure}
    \centering
	\includegraphics[width=0.95\columnwidth]{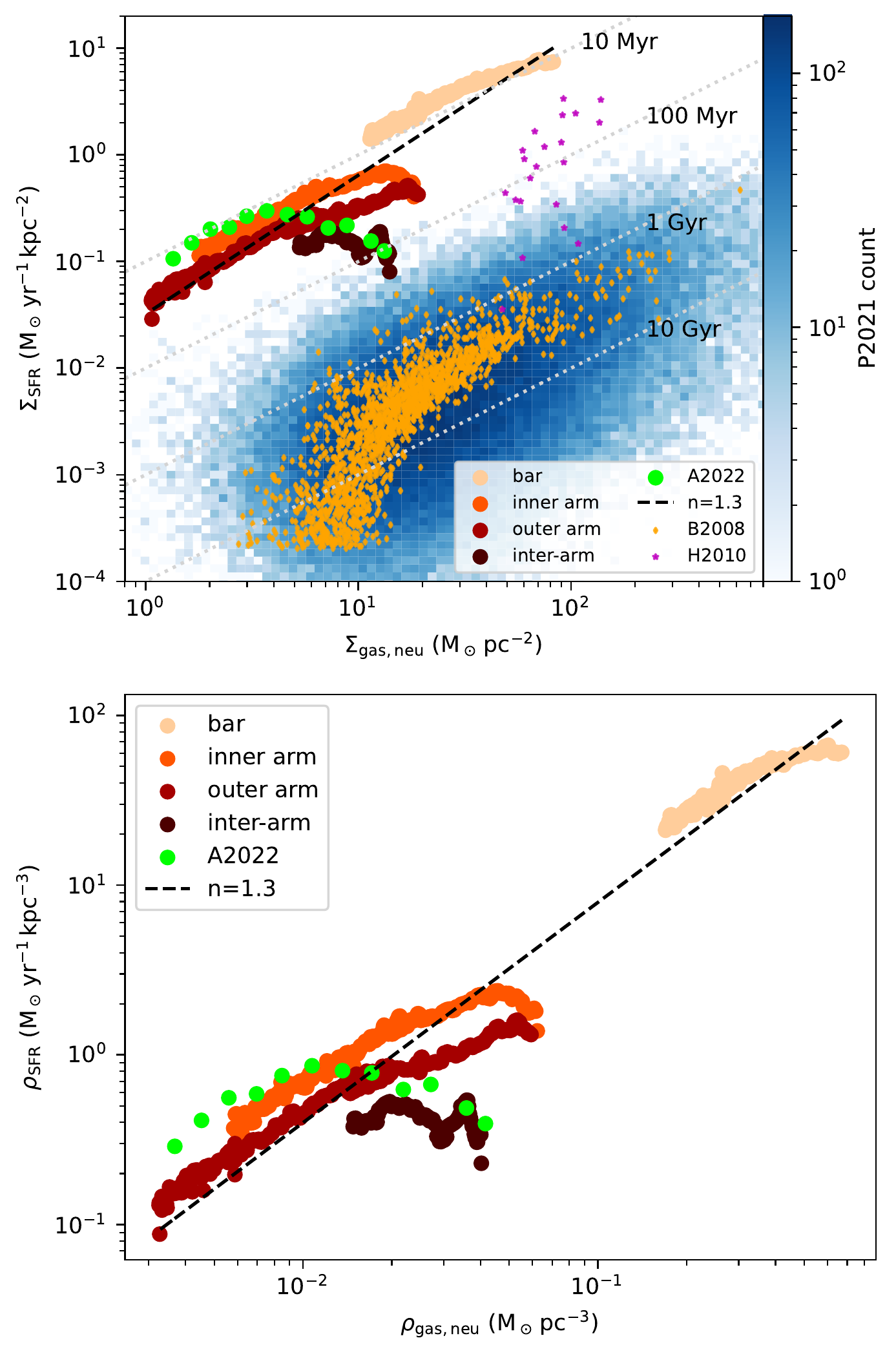}
    \caption{Top - Kennicutt-Schmidt relation between surface densities of the star formation rate vs. neutral gas mass (i.e. \ion{H}{i} + H$_2$). Bottom - the same but volume density instead of surface density. The green circles show results from \citet[][A2022]{ali2022}. The dashed line shows the power law fit with the index $n$ given in the legend. Dotted lines show gas depletion time scales between 10\,Myr and 10\,Gyr. Observations from \citet[][P2021]{pessa2021} are shown as a 2D histogram in colour scale. Data points from \citet[][B2008]{bigiel2008} are shown as orange diamonds and from \citet[][H2010]{heiderman2010} as magenta stars.}
    \label{fig:kennicutt}
\end{figure}

The top panel of \cref{fig:kennicutt} shows the relation between $\sigmasfr$ and $\sigmagas$, the gas mass surface density of neutral gas (equivalent to \ion{H}{i} + H$_2$). For each model, we plot a data point at every simulation dump time, starting from \SI{0.5}{\Myr} after the first ionizing source starts radiating; we exclude earlier points to avoid the initial large scatter in SFR before this time due to the small number of sinks (c.f. the middle panel of \cref{fig:sfe}). The black dashed line shows the power-law fit to the data, which is $\sigmasfr \propto \sigmagas^{1.3}$. This is found by applying a least-squares method to the data in log-space. The index agrees with the standard Kennicutt-Schmidt index of $1.4 \pm 0.15$ \citep{kennicutt2007}, especially at later times (lower $\sigmagas$) for the \bsda{}, \bsaa{} and \bsca{} models when the points evolve to follow the power law line. The \citet{ali2022} spiral arm zoom-in is also included here, and is shown in green. However, the \bsga{} model is shifted down compared to the other arm regions modelled in this paper -- at early times, $\sigmasfr$ is lower by a factor of 2--3. This model shows a double bump in SFR without tailing downward at later time (lower $\sigmagas$) like the arm and bar models; this is shown more clearly in \cref{fig:sfe} as a function of time. The bar region is located higher up the power-law line than the arm regions, which generally all have lower densities and lie at similar points in the figure.

For comparison, observed regions from \citet{bigiel2008} are shown as orange diamonds -- these are from 7 nearby spiral galaxies with \SI{750}{pc} resolution. Observations by \citet{pessa2021} from 18 galaxies at a resolution of 100\,pc are also plotted (assuming $\Sigma_\mathrm{mol,gas} \approx \Sigma_\mathrm{neu,gas}$). The models lie above the observational data -- our star formation rate surface densities are a factor of $\sim$100 higher. Lines of constant gas depletion time scale ($\propto \sigmagas$/$\sigmasfr$) are plotted as dotted lines. 
The models have gas depletion time scales between 10--\SI{30}{\mega\yr}, while the \citet{bigiel2008} and \citet{pessa2021} data show time scales of 1--\SI{10}{Gyr}, with some of the latter regions approaching 100-300\,Myr. Similarly, 1.5\,kpc resolution observations by \citet{sun2023} have depletion time scales above 1\,Gyr. Milky Way regions with higher resolution from \citet{heiderman2010} are shown in magenta. Their values of $\sigmasfr$ are closer to the simulations than the \citet{bigiel2008} results, but have higher $\sigmagas$ as they are individual star-forming clouds (sizes smaller than Orion) rather than cloud complexes. We test how the pixel size and sink parameters affect the results in \cref{sec:pixelsize}.

The bottom panel of \cref{fig:kennicutt} shows the star formation relation in terms of volume density (dividing by $XYZ$) instead of surface density -- this now includes the height of the region in the $z$-dimension as well (above/below the galactic plane). This is $\rhosfr$ against $\rhogas$. In observations, the main difference between the volume density law and the surface density law is that the latter is well known to have a break at low $\sigmagas$, while the volumetric law shows indications of being the same for all $\rhogas$ (for the same method). The simulations lie along the dashed line which shows the power law fit to the models, $\rhosfr \propto \rhogas^{1.3}$. The index of 1.3 agrees with one of the values calculated by \citet{bacchini2019,bacchini2019a}, who use surface density measurements in 12 nearby galaxies and turn this to a volumetric quantity by calculating the radius-dependent scale height of a disc in vertical hydrostatic equilibrium. They measure two power law indices, 1.3 (when using a constant SFR scale height of \SI{100}{pc}) and 1.9 (when using a radially varying SFR scale height).

\subsection{Cluster identification}
\label{sec:clusters}
We study the properties of the clusters formed in our simulations using two different methods, INDICATE \citep{buckner2019} and HDBSCAN \citep{hdbscan}. Both methods suggest the same trends in the properties of clusters with galactic region, namely that smaller denser clusters occur preferentially in the inner arm and bar regions, and larger clusters or associations in the outer arm and inter-arm regions. For this section, we analyse the clusters in each simulation at the time when the first supernova occurs (4.5\,Myr after the onset of feedback), except for the bar region which is analysed at 3\,Myr.

\subsubsection{INDICATE}
\begin{figure*}
   \centerline{
   \includegraphics[width=0.85\textwidth]{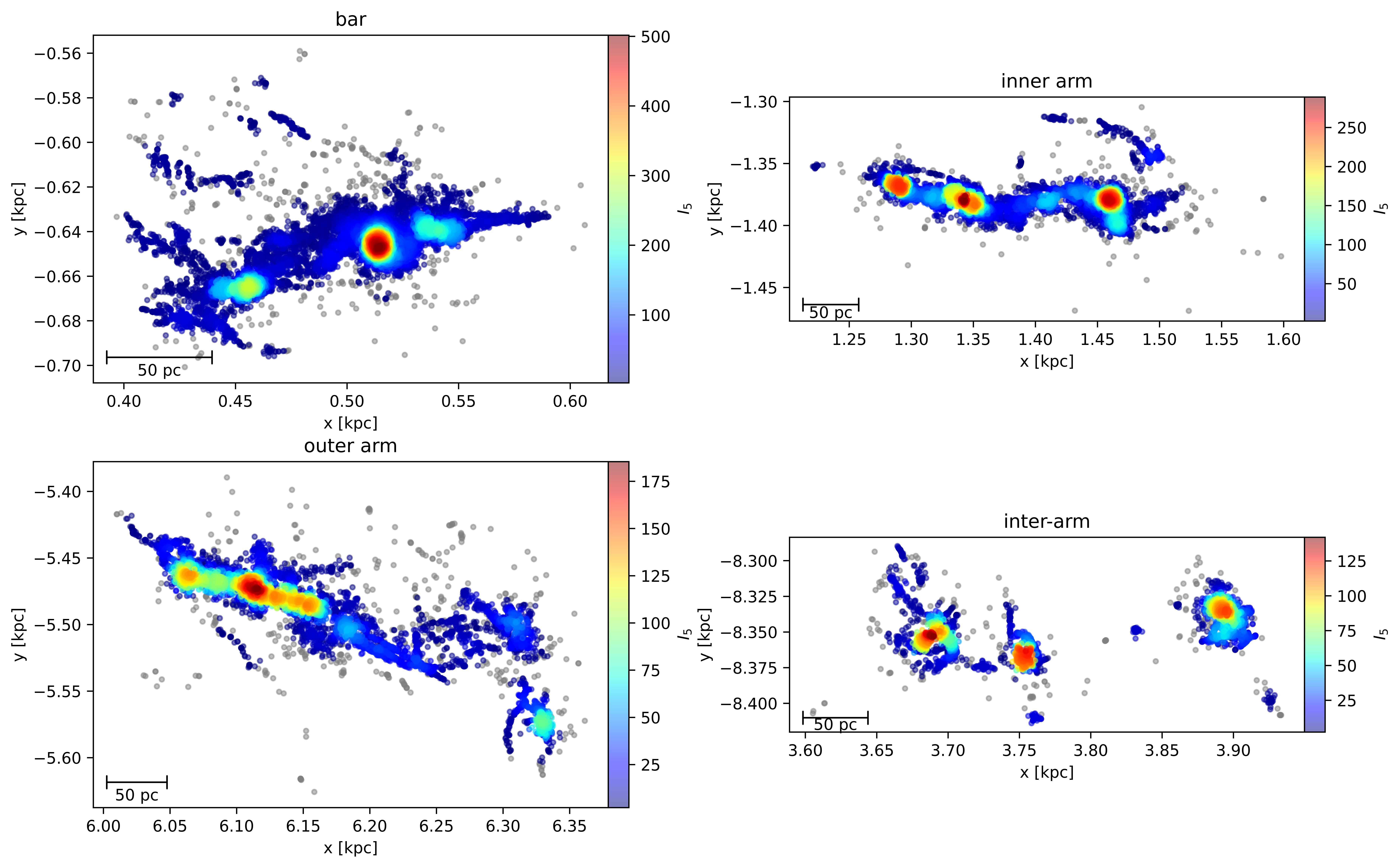}}
   \caption{Results from INDICATE for each model (top-down projection). The colour bar denotes the index value, $I_5$, of sinks found to be spatially clustered. Sinks with higher $I_5$ are more clustered. Points in grey denote sinks found to have a random distribution. Sinks in the bar region are the most clustered, followed by the inner arm, outer arm, and inter-arm regions.}
\label{fig:indicate}
\end{figure*}

INDICATE (INdex to Define Inherent Clustering And TEndencies; \citealt{buckner2019}) is a local indicator of spatial association. This means that rather than finding discrete clusters like (H)DBSCAN (see \cref{sec:hdbscan}), it assigns an index to each point in a dataset that describes the spatial distribution in its local neighbourhood. The index has a range of  $0\le\,I_5\le\frac{N_{\mathrm{tot}}-1}{5}$, where $N_\mathrm{tot}$ is the total number of points in the dataset and higher values represent greater degrees of association. INDICATE calibrates $I_5$ against random distributions to
identify the minimum value, $I_\mathrm{sig}$, which denotes a point is spatially clustered (rather than randomly distributed). This value is defined as three standard deviations greater than $\bar{I}_5^\mathrm{random}$ i.e. $I_\mathrm{sig}=\bar{I}_5^\mathrm{random}+3\sigma$. Statistical testing by the authors has shown INDICATE to be independent of dataset size, shape, and density; robust against edge effects and outliers; and valid for sample sizes $\ge50$ that are up to 83.3 per cent incomplete \citep{2022A&A...659A..72B}. 
INDICATE has been applied to observations of Carina \citep{buckner2019}, although care must be taken when comparing 2D projections with 3D data as the quantitative results may differ, while qualitative results are still consistent \citep{buckner2022}. Here, we use INDICATE to compare our four models with each other and provide a basis for identifying cluster members with HDBSCAN (\cref{sec:hdbscan}). These results may be useful for comparisons with future studies which use 3D \textit{Gaia} data.

We show the results of applying INDICATE to the different regions in \cref{fig:indicate}. The algorithm is applied to the 3D sink positions, with the figure showing the results projected onto the $x$-$y$ plane. The value of $I_5$ (colour scale) denotes the degree of association for sink particles that have been identified as spatially clustered, while randomly distributed sinks are plotted in grey. Sinks with higher values of $I_5$ are more clustered. \cref{fig:indicate} shows that the highest values occur in the bar, followed by the inner arm, the outer arm and inter-arm. For sinks identified as being clustered, the median value of $I_5$ in each model is 87 (bar), 92 (inner arm), 59 (outer arm), and 60 (inter-arm). Thus we see that denser, tighter clusters, are found in the bar and inner arm compared to the inter-arm and outer arm.

\subsubsection{(H)DBSCAN}
\label{sec:hdbscan}
\begin{figure*}
   \centerline{
   \includegraphics[width=0.85\textwidth]{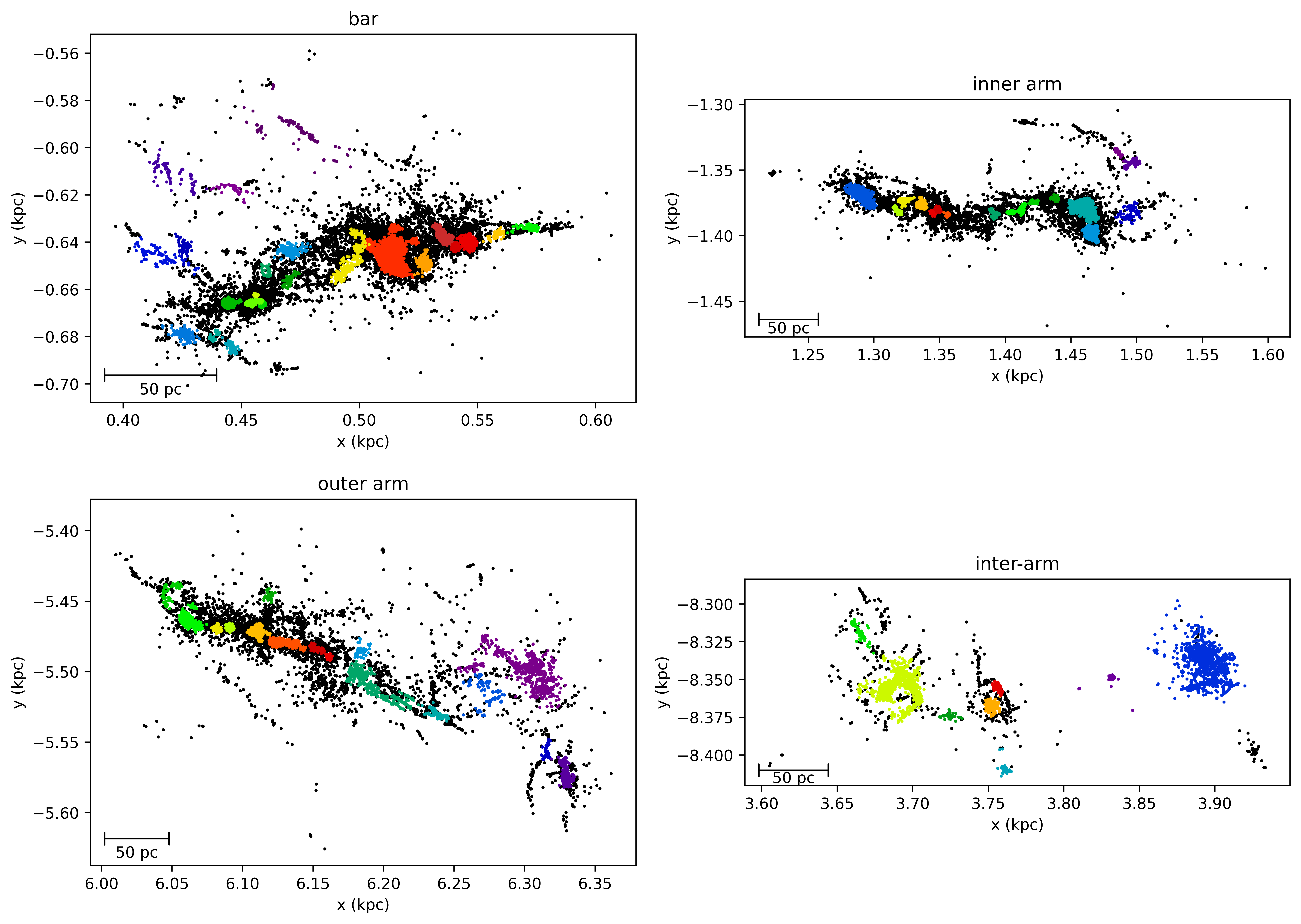}}
   \caption{Clusters of sink particles identified for each model using HDBSCAN (top-down projection). The different colours denote discrete clusters found, with members of the same cluster sharing colours. Particles which are not associated with any cluster are plotted in black. The clusters tend to be more spatially extended in the outer arm and inter-arm models, compared to the bar and inner arm models.}
\label{fig:clusters_xy}
\end{figure*}

DBSCAN \citep{ester1996} is an algorithm commonly adopted observationally to identify clusters (\citealt{2019A&A...627A..35C}, \citealt{2020A&A...635A..45C}, \citealt{2022Univ....8..111C}, \citealt{2022A&A...664A.175P}, \citealt{2022ApJS..262....7H},  \citealt{2022A&A...661A.118C}, \citealt{2023ApJS..264....8H}). It requires two input parameters, $\epsilon$ and \texttt{MinPts} and categorises stars as either the core/border members of a cluster, or noise. $\epsilon$ is the radius around each star $j$ that is searched for neighbouring stars, and \texttt{MinPts}-1 is the minimum number of neighbours needed for $j$ to be a core member of a cluster. If $j$ has less than this minimum number of neighbours but is within the search radius of a core star it is considered a border star, else noise. We chose \texttt{MinPts}$=$30 simply as measure of the size of a cluster which is well resolved in the simulation. We originally tried using DBSCAN to identify clusters, ideally keeping \texttt{MinPts} and $\epsilon$ the same so that clusters could be compared consistently across different models, but we could not find uniform values of the $\epsilon$ parameter across all models. Using the method proposed by \citet{ester1996} to determine $\epsilon$, with \texttt{MinPts}$=$30, gave values of 0.39, 0.55, 0.67 and 0.88\,pc for the bar, inner arm, inter-arm region and outer arm respectively, indicating that the bar and inner arm contain smaller clusters compared to the other regions.  With these values of $\epsilon$, the bar and inner arm preferentially produce smaller, denser clusters whereas the outer arm and inter-arm regions produce spatially larger, lower density clusters.

Although DBSCAN is commonly used in the observational literature, a disadvantage is that it is designed to find clusters of similar densities. This is especially true for datasets such as ours, in which clusters of different densities exist side-by-side in the same region, and different regions have different densities as well. HDBSCAN \citep{hdbscan} is a successor to DBSCAN which utilises a hierarchical clustering approach to find clusters in different density regions. This increased sensitivity makes HDBSCAN a more effective algorithm for recovering clusters in observational datasets \citep{2021A&A...646A.104H}. HDBSCAN does not require the $\epsilon$ parameter to be user-defined as it is essentially an implementation of DBSCAN which varies this value. Instead the main input to HDBSCAN is \texttt{min\_cluster\_size}, indicative of the minimum number of points in a cluster. A second input is \texttt{min\_samples}, which is a measure of how strict the cluster assignment is. 
We found HDBSCAN produced satisfactory clusters without the need to alter the $\epsilon$ parameter, and unlike DBSCAN, we use the same parameters for HDBSCAN to find clusters across all the models. 

To identify the best choice of input parameters for HDBSCAN, we determined the median INDICATE values of the clusters identified across the different models, giving us a range of values for \texttt{min\_cluster\_size} and \texttt{min\_samples} which produced similar $I_5$ values (note that there is some degeneracy between the two values, so for example increasing \texttt{min\_cluster\_size} and \texttt{min\_samples} both tend to produce larger clusters). We also checked the clusters identified with the different parameters by eye, and compared the peaks identified with INDICATE (see \cref{fig:indicate}) with the resultant clusters. We found that the inner arm, inter-arm, and outer arm favoured smaller values of \texttt{min\_cluster\_size} and \texttt{min\_samples}, whilst the bar favours larger values. This again indicates that the bar model produces denser clusters with more particles (similar to the INDICATE results).
We selected the largest values of \texttt{min\_cluster\_size} and \texttt{min\_samples} within our optimal range which did not spuriously group particles together which were not clusters by eye in the inter-arm, inner arm and outer arm models. These values then overlapped with the lower range of optimal values for the bar model. Overall this approach gave values of \texttt{min\_cluster\_size}$=55$ and \texttt{min\_samples}$=40$. 

We show the clusters picked out with the HDBSCAN algorithm using these parameters for the different models in \cref{fig:clusters_xy}. As with INDICATE, HDBSCAN uses the 3D sink positions, with the figure showing a 2D projection. Although the scales vary slightly between the different panels, the figure indicates that more spatially extended clusters are found in the inter-arm model, and to some extent the outer arm model, whereas more compact clusters are found in the inner arm and bar models.

\subsection{Cluster masses and radii}
\label{sec:clustermass}

\begin{figure}
   \centerline{\includegraphics[width=0.95\columnwidth]{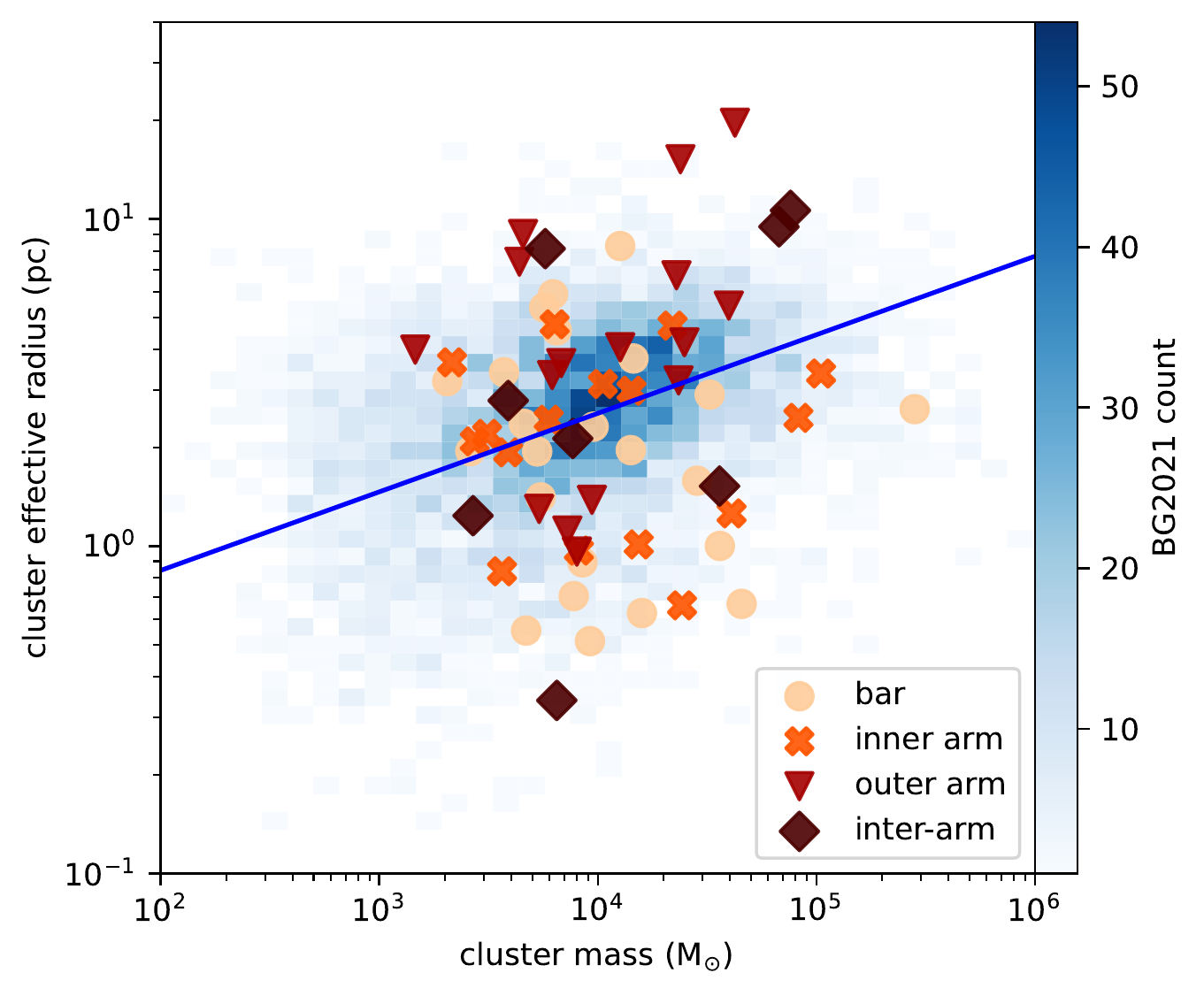}}
   \centerline{\includegraphics[width=0.95\columnwidth]{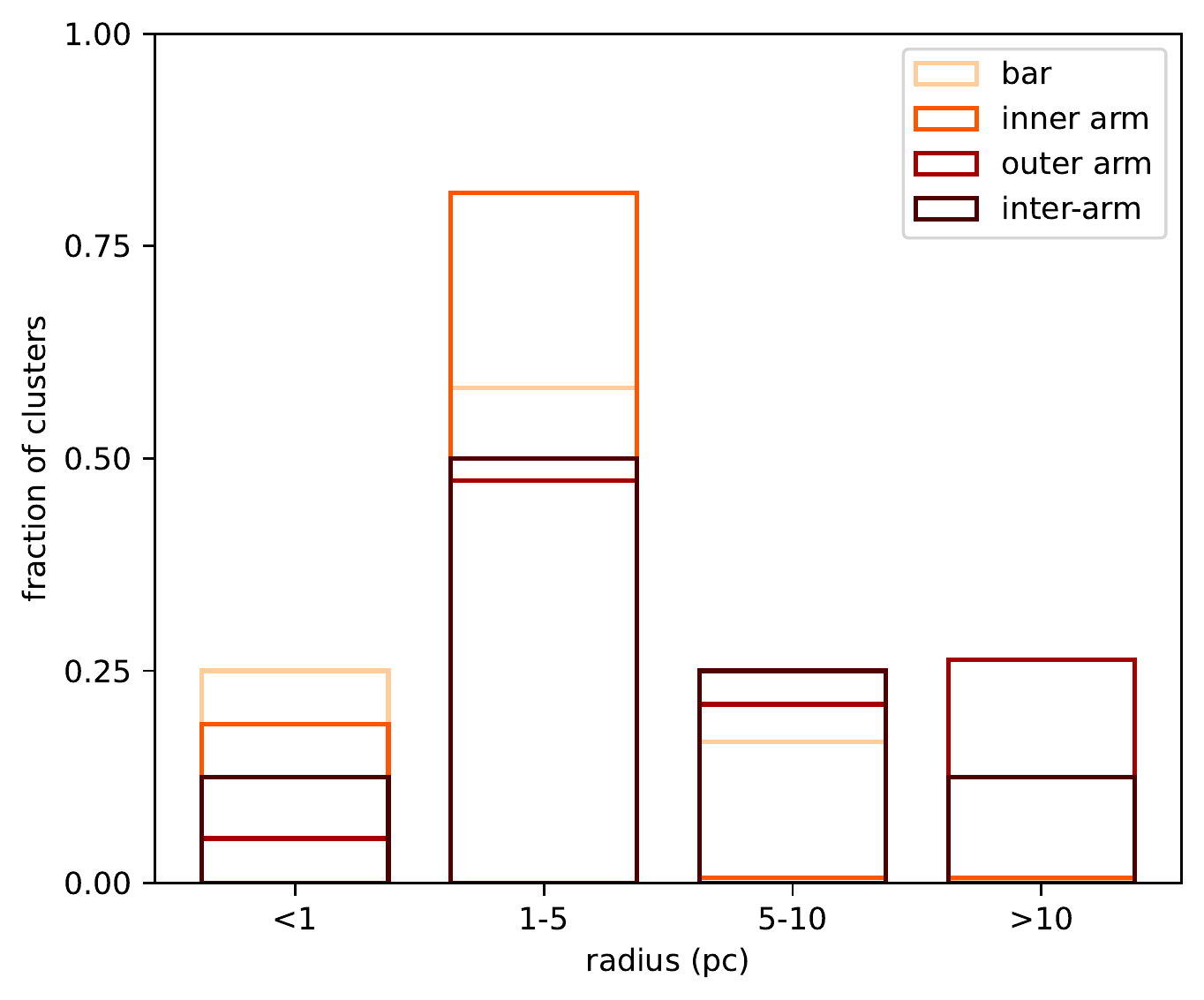}}
   \caption{Top - cluster radius vs. mass for the different models, plotted over observational data from \citet{brown2021} as a 2D histogram. The solid line shows the fit to the observations. Bottom - fraction of clusters in different radius bins. Clusters with radii $>$ 5\,pc are predominantly found in the outer arm and inter-arm regions, whilst no spatially larger clusters are found in the inner arm region.}
\label{fig:clusters_rad}
\end{figure}

Having identified the clusters, we show their effective radii and masses in the upper panel of \cref{fig:clusters_rad}, where the data from our models is plotted over observational data from \citet{brown2021}. The effective radius we use is the half-mass radius, which is comparable to the observed radius. Similarly to \citet{dobbs2022a}, the points collectively indicate very similar trends to the observed data, and a similar increase in radius with mass. As also found in \citet{dobbs2022a} the points have a slightly larger spread compared to the observational data. As mentioned in \cref{sec:sfr}, the star formation rates are higher than would be expected, so the clusters tend to be at the high mass end of the observational data. We find that the clusters with the largest radii form in the inter-arm and outer arm models -- this is also indicated by eye in \cref{fig:clusters_xy}. With radii of around 10 pc, these objects are more comparable to local observed associations \citep{portegies-zwart2010} than clusters (note that smaller associations may be part of much larger $\sim$100 pc regions or associations \citep{Wright2020} but here we compare with the $\sim$ 5 pc size associations discussed in \citet{portegies-zwart2010}). The clusters in the bar and inner arm models appear to follow fairly well the observed distribution. The clusters in the outer arm however exhibit relatively large radii, and taken in isolation would exhibit a much steeper radius mass relation compared to the observational dataset. 

In the lower panel of \cref{fig:clusters_rad}, we plot the frequency of clusters with different radii for the different models. Again the bar and inner arm models contain spatially smaller clusters compared to the outer arm and inter-arm regions. The inner arm in particular contains no clusters of radii $>5$ pc, whereas nearly half the clusters in the outer arm, and over a third in the inter-arm region, have radii $>5$ pc.

We note that the tendency of the spatially largest clusters, for a given mass, to occur in the outer arm and inter-arm models is independent of our choice of algorithm or input parameters for HDBSCAN. Using DBSCAN with the above $\epsilon$ values (see \cref{sec:hdbscan}), or choosing lower values for \texttt{min\_cluster\_size} and \texttt{min\_samples} with HDBSCAN, tends to break up the clusters more, and the points are shifted to lower masses; in which case the larger clusters for the outer arm and inter-arm models are shifted to the top left area of \cref{fig:clusters_rad} (upper panel).

We also see from \cref{fig:clusters_rad} (upper panel) that the most massive cluster is formed in the bar, then the second most massive in the inner arm, followed by the inter-arm region, then the outer arm. Again this trend is fairly robust to the choice of algorithm and input parameters. We can compare the cluster masses and radii with clusters in the Milky Way. Although there is not a complete map of clusters in our Galaxy, \citet{portegies-zwart2010} list the main clusters (or YMCs) and associations in our quadrant. At the end of the bar lies RSGC02 which is the most massive cluster (using $M_\mathrm{phot}$ from table 2 of \citet{portegies-zwart2010}), at $4\times10^4$\,M$_{\odot}$ (and RSGC01 and RSGC03 are close by with similar masses). The next most massive is Westerlund 1, in an inner spiral arm. Between arms, on a minor spiral arm, lies Orion which according to the table has a combined mass of $2\times10^4$\,M$_{\odot}$, whilst NGC 3603 just outside the solar circle has a mass of $1.2\times10^4$\,M$_{\odot}$. Therefore the trend in cluster mass with region seen in the simulations is the same as seen in our Galaxy.

\subsection{Cluster rotation and expansion}

\begin{figure}
    \centering
    \includegraphics[width=0.95\columnwidth]{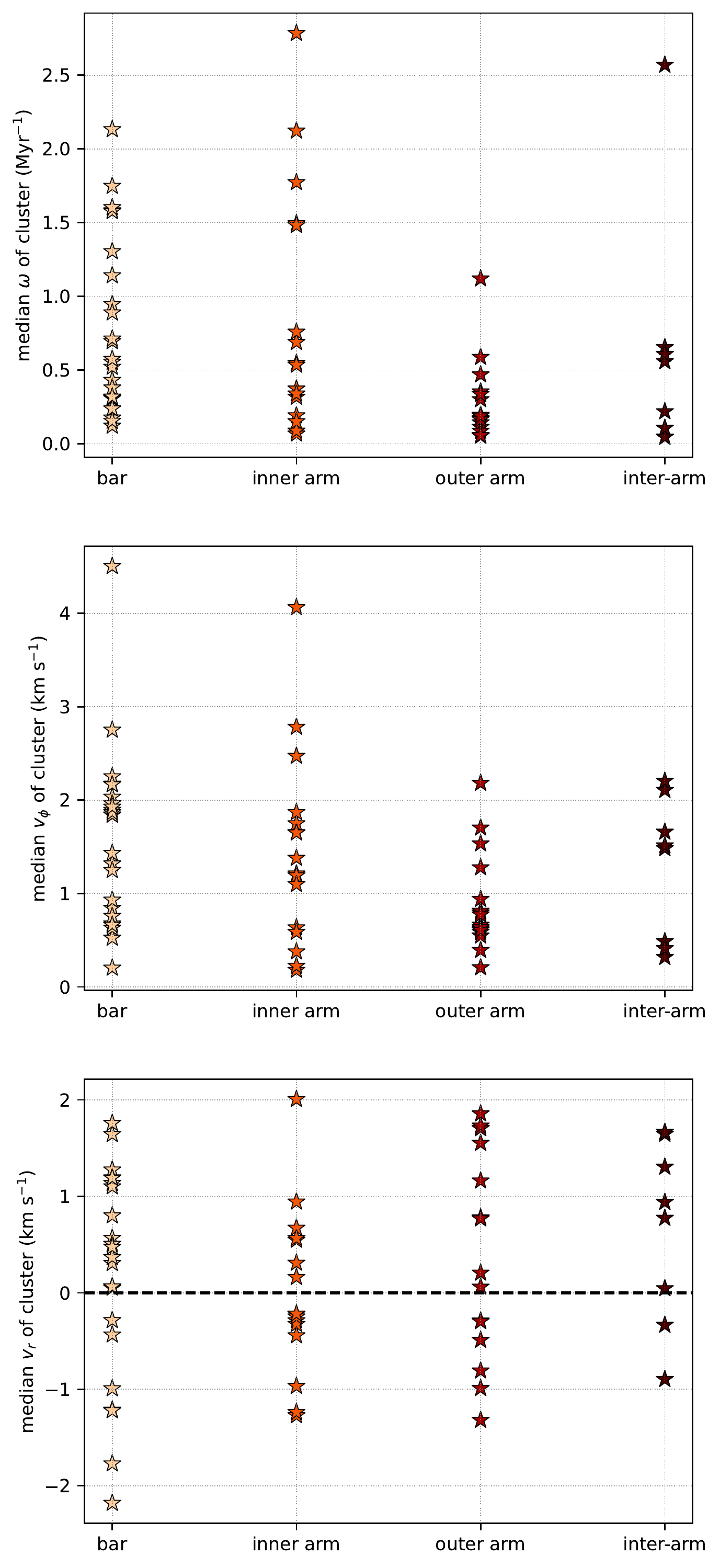}
    \caption{Median values of the angular velocity $\omega$ (top), azimuthal velocity $v_\phi$ (middle), and radial velocity $v_r$ (bottom) for each cluster identified in \cref{fig:clusters_xy}. Positive/negative $v_r$ indicates radial expansion/contraction respectively.} 
    \label{fig:rotation}
\end{figure}

We measure the bulk rotation of the clusters identified in \cref{sec:clusters}. We use a similar method to \citet{ballone2020} and \citet{verliat2022}. For each cluster, we identify the centre of (sink) mass and calculate the angular momentum of each sink $\mathbfit{L} = m \mathbfit{r} \times \mathbfit{v}$, where $\mathbfit{r}$ and $\mathbfit{v}$ are its position and velocity, respectively, relative to the centre of mass. We then calculate the mean angular momentum $\mean{\mathbfit{L}}$ and rotate the reference frame so that the new $z'$ axis is parallel to $\mean{\mathbfit{L}}$, meaning this is defined to be the rotation axis of the cluster. For each sink, we calculate the azimuthal velocity component $v_\phi = \mathbfit{v}' \cdot \hat{\mathbf{\phi}'}$, where $\hat{\mathbf{\phi}'}$ is the azimuthal unit vector around the $z'$ axis. The angular velocity is then $\omega = v_\phi / \varrho$, where $\varrho$ is the distance from the $z'$-axis. The radial velocity component is $v_r = \mathbfit{v}' \cdot \hat{\mathbfit{r}'}$. 

In \cref{fig:rotation}, we plot the median values of $\omega$ (top panel), $v_\phi$ (middle panel), and $v_r$ (bottom panel). All the clusters rotate with a non-zero median velocity. The bar and inner arm clusters generally have higher angular velocities than the outer arm and inter-arm regions. The clusters in the outer arm and inter-arm models  with median $\omega > \SI{0.75}{\per\Myr}$ appear to be outliers compared to the rest of the clusters in those regions, whereas the bar and inner arm have several clusters above this value, even extending beyond \SI{2}{\per\Myr}.  The median of medians for the angular velocities $\omega$ in \si{\per\Myr} are 0.57 (bar), 0.54 (inner arm), 0.18 (outer arm), and 0.22 (inter-arm) excluding the largest outlier.  Similarly, the median of medians of $v_\phi$ in \si{\kms} are 1.8 (bar), 1.3 (inner arm), 0.72 (outer arm), 1.5 (inter-arm). Note that while clusters in the inter-arm may have higher $v_\phi$ on average compared to the inner arm, they are also larger in size (see \cref{fig:clusters_rad}), hence the division by $\varrho$ results in a smaller $\omega$. The median radial velocities show that the majority of clusters in each region are expanding, except in the inner arm where most (56 per cent) are contracting. Radial velocities rarely exceed $\pm \SI{2}{\kms}$, which is consistent with the observed clusters listed in table 2 of \citet{kuhn2019}. The only cluster with statistically significant rotation in the \citet{kuhn2019} dataset is Tr 15 in the Carina Nebula, which lies in a spiral arm about \SI{2.4}{kpc} from the Sun \citep{shull2021}. It has a median $v_\phi = \SI{1.7 \pm 0.5}{\kms}$, which is consistent with the velocities we find in our models, but less so with the outer arm region where most of the clusters rotate more slowly than this. R136 in the Large Magellanic Cloud has a mean $v_\phi = \SI{3 \pm 1}{\kms}$ and $\omega = \SI{0.75 \pm 0.22}{\per\Myr}$ \citep{henault-brunet2012}, which fits with the bar/inner arm values, but is on the extreme end of the outer arm/inter-arm results.

\subsection{Mass flows}
\begin{figure*}
    \centering
    \includegraphics[width=0.95\textwidth]{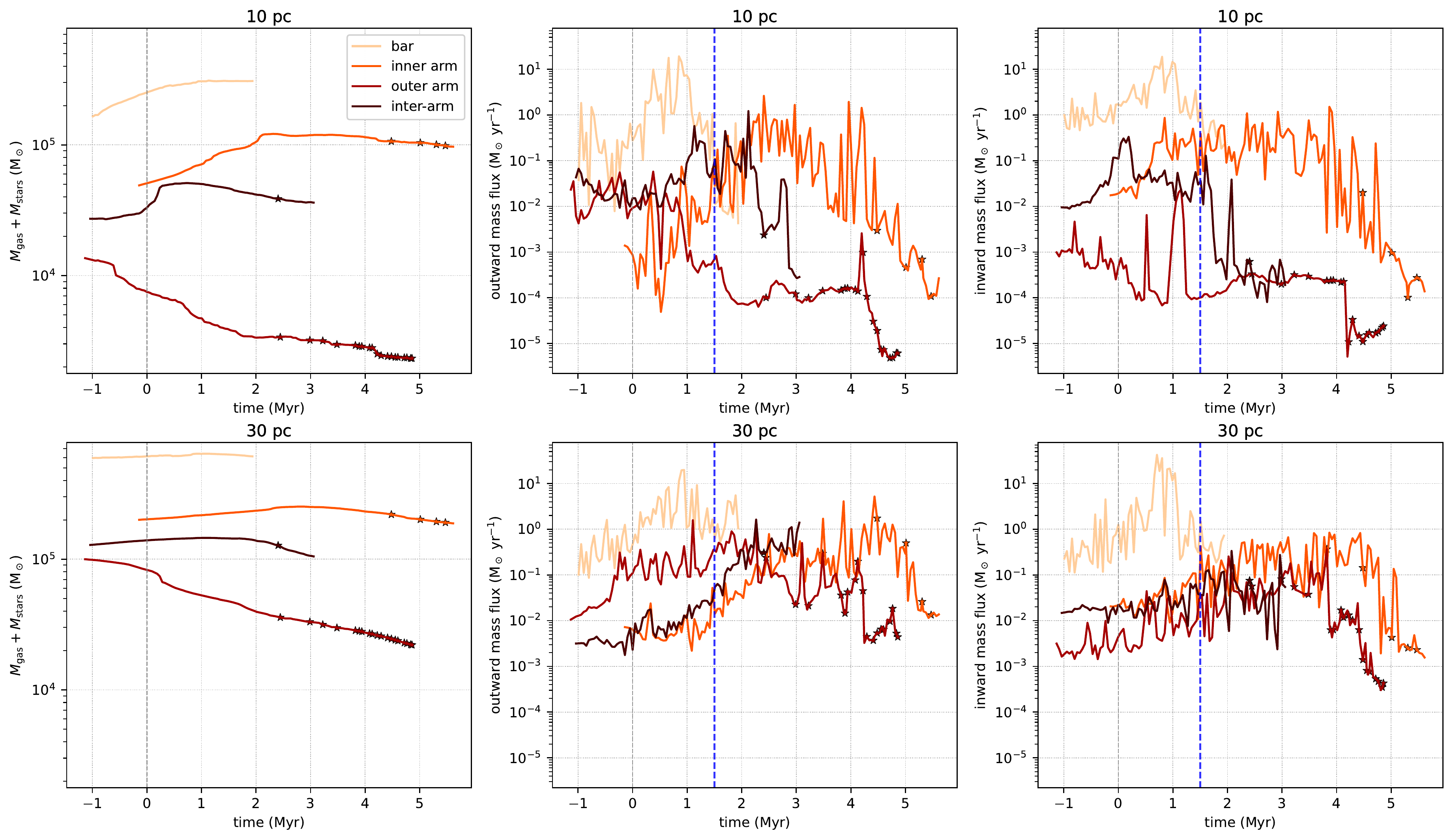}
    \caption{Total mass inside, and mass flux through, a spherical surface at two radii: \SI{10}{pc} (top row) and \SI{30}{pc} (bottom row). The left column shows the total mass in the sphere. The mass flux is separated into the radially outward direction ($\dot{M}_+$, middle column) and the radially inward direction ($\dot{M}_-$, right column). Zero on the time axis is when the origin sink starts emitting ionizing radiation. The lines start when the sink is first formed. See \cref{fig:massfluxcolden} for column density snapshots at the time highlighted by the blue dashed line. Star symbols denote supernova events.} 
    \label{fig:massflux}
\end{figure*}

\begin{figure*}
    \centering
    \includegraphics[width=0.95\textwidth]{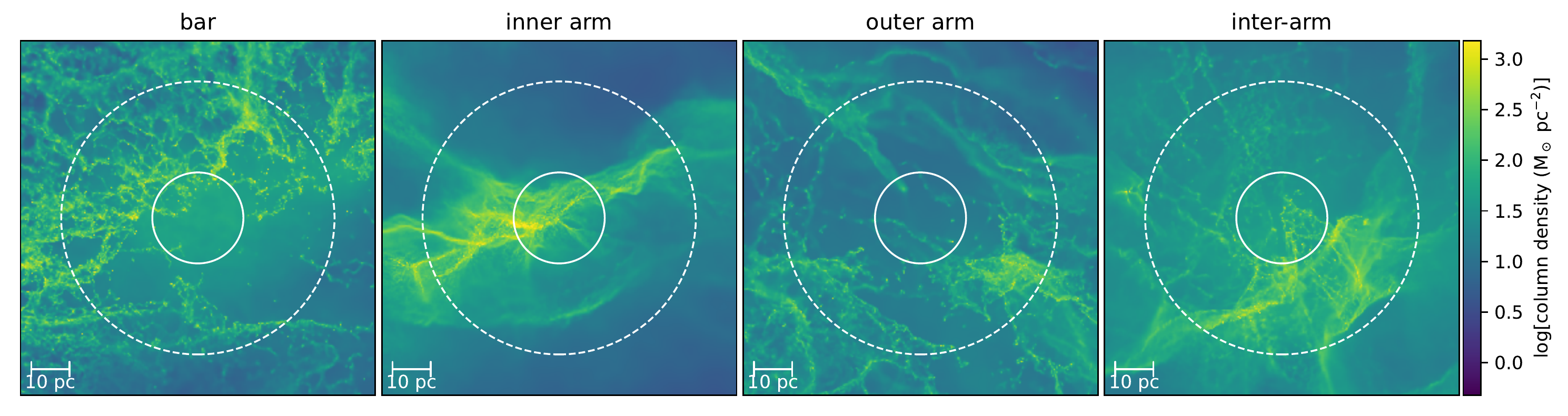}
    \caption{Column density snapshots  ($x$-$y$ plane) in the sub-region investigated in \cref{fig:massflux} at the time highlighted with a blue dashed line. Mass fluxes are calculated at \SI{10}{pc} (inner circle) and \SI{30}{pc} (outer circle) from the origin sink. } 
    \label{fig:massfluxcolden}
\end{figure*}
We investigate the dispersal or replenishment of gas under the influence of feedback by measuring the radial mass flux near clusters. First, we identify the cluster with the highest total ionizing flux, and then the sink in that cluster with the highest ionizing flux -- we do this at the same snapshots as in \cref{sec:clusters}, just before the first supernova for the inner arm, outer arm, and inter-arm models (4.5\,Myr after feedback starts), and at 3\,Myr for the bar region. We define the position of this sink as the origin of a sphere of radius $R$ and locate particles at the surface, for which we calculate the instantaneous mass flux through the surface (radially outward and radially inward). To calculate this numerically, we discretize the surface using HEALPix \citep{gorski2005}, creating equal-area cells ($\Delta{S}$) at a defined radius. We locate particles between $R$ and $(R-\SI{1}{\pc})$, and sort them according to HEALPix cell. In each cell, we calculate the mean value of the mass flux per unit area $\rho v_r$, where $\rho$ is mass volume density and $v_r = \mathbfit{v} \cdot \hat{\mathbfit{r}}$, i.e. the  velocity (relative to the origin sink) in the direction pointing radially away from the origin sink. We then integrate this value over cells to calculate the mass flux in units of \si{\msol\per\yr},
\begin{equation}
\label{eq:massflux}
    \dot{M} = \sum_{\mathrm{cells}~i} \mean{\rho v_r}_i \Delta{S_i}~,
\end{equation}
separating out the positive (outward) components and the negative (inward) components. We do this for two radii, $R$ = \SI{10}{pc} and \SI{30}{pc} representing small and large scales respectively. For reference, if gas moves radially outward from the origin at the ionized sound speed ($\sim \SI{10}{\kms}$), it will take \SI{1}{\mega\yr} to reach \SI{10}{pc} and \SI{3}{\mega\yr} to reach \SI{30}{pc}. We track the origin sink backwards through time and repeat the procedure every \SI{0.047}{\mega\yr}. 

In \cref{fig:massflux}, we plot the total mass inside the sphere as a function of time. We also plot the mass flux flowing radially outward through the surface, $\dot{M}_+$, and inward, $\dot{M}_-$. For context, \cref{fig:massfluxcolden} shows column densities at the time highlighted by the blue dashed line of \cref{fig:massflux} -- this shows the differences in gas morphology across the region at 1.5\,Myr, enough time for radiation to propagate and ionize gas, and for gas to reach a surface.

While feedback is occurring, the \bsaa{} and \bsca{} models shows the most significant change in total mass inside 10\,pc (top row, leftmost panel). In the \bsaa{}, the mass increases by more than a factor of 2, from 0.5 to \SI{1.2e5}{\msol} at 2.3\,Myr. The streams of dense gas are more collimated compared with the other regions where dense gas is spread out more evenly over the surfaces.  After the peak mass is reached, the total mass then declines slowly for more than 3\,Myr as gas is gradually dispersed from the system by ionization (while most of the sinks actually fall into the 10\,pc sphere over this time). 

On the other hand, the \bsca{} shows a decline in mass over the whole runtime, decreasing by a factor of 2 within the first 2\,Myr of feedback. This is partly due to the gas dynamics set by the original galaxy simulation, which shows up as outflow when clouds move across the region, separate to the effect of stellar feedback. As the sinks form and feedback progresses (from this origin sink as well as nearby sinks), the decrease is due to the sinks dispersing in addition to the gas itself. This model is already the lowest density of the four sub-regions (\cref{fig:massfluxcolden}), meaning feedback does not have to be strong to disperse it.  The \bsca{} has low inward fluxes over the whole runtime, but does exhibit large spikes as clumps of material pass through the surface; however, these stop after about 1.3\,Myr as much of the material has left by this point -- the outward mass flux declines from \SI{e-2}{\msol\per\yr} around 1\,Myr to \SI{e-4}{\msol\per\yr} by 2\,Myr. 

The \bsga{} follows a similar pattern as the \bsaa{} despite the different morphology, albeit over a shorter timespan. It increases its mass by a factor of 2 by the time the feedback starts, at which point this plateaus and the total mass slowly declines (c.f. the \bsaa{} where this takes 2\,Myr). This region has high outward mass fluxes (top row, middle panel), with values often exceeding \SI{e-1}{\msol\per\yr} and becomes the model with the second-highest flux (behind the bar). The inward mass flux (right panel) drops by almost two orders of magnitude around 1.5\,Myr, meaning this region disperses material effectively while not maintaining any infall -- the infall here then matches the \bsca{}, reaching the smallest values of a few \SI{e-4}{\msol\per\yr}.  

The \bsda{} shows the smallest change to the total mass inside 10\,pc, simply increasing its mass slowly over the course of the simulation. The material in this model rotates around, causing mass to enter the surface and a comparable amount of mass to leave it, with  high mass fluxes of 1--\SI{10}{\msol\per\yr} being reached in this model. Indeed, in the first 1.5\, Myr at 10\,pc, the \bsda{} has the highest inward mass fluxes, followed by the \bsaa{}, \bsga{}, then \bsca{}. 

At 30\,pc (bottom row), the inward fluxes are generally similar for all models except the \bsda{}, which is 1--2 orders of magnitude higher than the other models, especially in the first 1.5\,Myr. The other models are all similar to each other, with any morphological differences averaging out over this larger radius. On the other hand, the outward mass fluxes differ more strongly between the models, with the \bsca{} being an order of magnitude greater than the \bsaa{} and \bsga{} in the first Myr. The fluxes for the latter two slowly increase to match the \bsca{} by 2--2.5\,Myr, as the dispersive effect of feedback increases in these regions (which are initially denser than the \bsca{}). Overall, the models differ from each other more noticeably when considering the local gas flows (10\,pc), with less difference between regions at larger distances (30\,pc); the exception is the bar, which has higher fluxes than the other regions even at the larger radius.

\section{Discussion}
\label{sec:discussion}

There have now been multiple studies showing the variation of giant molecular cloud (GMC) properties with environment, initially with individual galaxies \citep{Colombo2014, Pan2017}, but now across larger samples with the PHANGS survey \citep{Sun2018}. The variation of cluster properties is less established, but there is some observational evidence of variation with environment. \citet{Messa2018} find that the cluster mass function is steeper in the interarm region of M51, whilst higher mass clusters are present in the inner parts of M83  \citep{Adamo2015,della-bruna2022a}. In our models, we also find that the more massive clusters form in the inner regions compared to outer regions, and also in spiral arms compared to inter-arm regions. Our cluster masses also vary with environment similarly to the Milky Way. 

We see that the bar region produces the most massive and dense clusters, and the star formation rate is highest here. Observationally, there are often particularly massive clusters at the ends of bars. For example in the Milky Way, the possible super-star cluster W43 appears to be forming at the end of the bar \citep{nguyen-luong2011,carlhoff2013}. 
High density gas can be collected together at the bar ends \citep[as seen in the galaxy NGC 3627;][]{beuther2017}, and strong tidal fields and turbulence can create higher density clouds in the Galactic Centre generally compared to the disc \citep{oka2001,henshaw2016,kruijssen2019} -- these extreme initial conditions could produce high mass clusters. 
Massive clusters are observed along the dust lanes of the disc-shaped region surrounding the bar in NGC 1365 \citep{Elmegreen2009,Schinnerer2023}. This is similar in morphology (although larger) to the ring structure produced by the bar in our galaxy simulation.  This seems to contrast with some surveys \citep[e.g.][]{Sheth2000,Momose2010}, which find that there is a lower star formation rate in the bar. However, if we consider the wider bar region of the original galaxy simulation (\cref{fig:initialinset}, first panel), we see that there are also large areas with very little gas. So the wider bar region contains both low density areas and very dense ring or disc-like structures (see also \citealt{Renaud2015,Shimizu2019,Querejeta2021,Iles2022,Maeda2023}). 

Generally, the trends we see with environment are equivalently trends with initial density, but it is not necessarily possible to readily distinguish between the two. For example, as we suggest above, the particular dynamics of the bar lead to very high densities, and similarly the spiral arms are regions of more strongly convergent flows. 
We do see some indication of a difference in star formation rate surface density between the arm and inter-arm region for a given gas surface density. The spiral arms may gather gas together, forming large complexes with increased star formation rates compared to other clouds \citep{dobbs2017}, though it is not clear whether the arms make a large difference to the galactic star formation rate \citep{dobbs2011,eden2013,eden2015,pettitt2020b,Urquhart2021}.

We find that our star formation rates are high compared with observed extragalactic regions (\cref{fig:kennicutt}). One explanation for the discrepancy between our results and the \citet{bigiel2008} observations is the latter have coarser spatial resolution -- see also \cref{sec:pixelsize} and the discussion in \citet{heiderman2010}, comparing the extragalactic results to Milky Way regions and discussing the effect of pixel size. This means regions where little or no star formation is taking place is also included in the observed SFR measure, driving down the spatially averaged $\sigmasfr$. The simulated boxes, meanwhile, encompass the star forming regions without including too much of the non-star forming material -- and indeed, there is a selection bias to the measurements as the initial conditions were chosen in part for their potential to form stars.  
This does not totally explain the discrepancy, however, as our star formation rates/efficiencies are still higher than most of the 100\,pc resolution data from e.g. \citet{chevance2020}, \citet{pessa2021}, and \citet{kim2022a}.
Given the computational difficulty of resolving star formation self-consistently on these scales, it is necessary to use a sub-grid model instead. Together, our tests in \cref{sec:pixelsize} imply the input parameter for the star formation efficiency per cluster-sink is too high (currently 50 per cent) and that the sink accretion radii may need reducing to avoid too much material being accreted. It is also possible that our star formation method is triggered at lower densities than they should be, as our simulation results are slightly shifted to the left (lower gas density) compared to resolved Milky Way clouds. Our clouds can still be compared reliably to each other as they have the same simulation parameters, and in terms of observations, represent clouds with active, high star formation.

Measurements of the rotation of young clusters are rare \citep{kuhn2019}, but our results predict that rotation is highest in the bar and inner arm, which could simply reflect the higher angular velocity of the galaxy at smaller radii. Finally, our simulations also suggest that differences in the gas flux from the wider environment ($>$30 pc) are minor -- however, there are larger differences between gas inflows/outflows to/from clusters on 10 pc scales. The exception is the bar area, with gas inflow over larger scales in this region, again likely because of the high rotation of the bar compared to the arms. The inflow rates for the bar region are comparable to the values observed flowing along the dust lanes towards the Central Molecular Zone of the Milky Way \citep{sormani2019} and the nuclear ring of NGC 1097 \citep{sormani2023}. 

%
%
\section{Summary and conclusions}
\label{sec:conclusions}

We present zoom-in simulations of cloud complexes extracted from a galaxy evolution model similar to the Milky Way, which contains a bar and four spiral arms \citep{pettitt2020}. Clouds have been taken from the bar, inner and outer spiral arms, and an inter-arm region, with masses \SI{2e6}{\msol} and sizes 100--\SI{300}{\pc}. The zoom-ins include ray-traced photoionization from cluster-sink particles \citep{bending2020}. The new resolution is \SI{0.4}{\msol} per particle, compared to \SI{600}{\msol} per particle in the original galaxy run. We have calculated star formation measures and cluster properties as a function of galactic environment. Clusters have been identified with HDBSCAN \citep{hdbscan} and the degree of clustering measured with INDICATE \citep{buckner2019}. Our key results are:
\begin{enumerate}
\item Denser regions form stars at a higher rate, following the relation $\sigmasfr \propto \sigmagas^{1.3}$, which is consistent with the Kennicutt-Schmidt index \citep{kennicutt2007}. However, the inter-arm model forms stars less efficiently than the spiral arm regions for the same $\sigmagas$, as $\sigmasfr$ is a factor of 2--3 below the arms. The bar is always the most star-forming model.
\item Almost all the clusters in the bar and inner arm are smaller than $\SI{5}{pc}$. Half the clusters in the outer arm and a third in the inter-arm are larger than $\SI{5}{pc}$, with radii more similar to associations.
\item Similarly, applying INDICATE shows that the degree of clustering is highest in the bar and decreases sequentially down to the inter-arm.
\item The bar and inner arm regions are able to form faster rotating clusters, while the outer arm and inter-arm regions tend to produce slower rotators on average. The representative angular velocities $\omega/\si{\per\Myr} =$ 0.57 (bar), 0.54 (inner arm), 0.18 (outer arm), and 0.22 (inter-arm).
\item The dispersive effect of feedback is shown through radially outward mass fluxes measured at spherical surfaces around the most ionizing cluster. Gas streams away from massive stars, and dense clumps show up as bursts of high fluxes. Radially inward fluxes can still be maintained for the bar and inner arm. Regions differ from each other the most at the smaller scale (10\,pc), whereas they are more similar at the larger scale (30\,pc). 
\end{enumerate}

These models do not include stellar winds. In previous zoom-in models of a different galaxy, we have shown this can affect the sink and cluster properties, typically producing smaller clusters \citep{ali2022}. However, winds only have a minor effect on the gas dynamics and morphology compared to photoionization \citep[see also][]{gatto2017,rathjen2021}. Finally, we do not discuss the role of supernovae as the bar region has not yet evolved to the point of the first supernova, making a comparison between regions difficult. We expect to investigate the role of supernovae in different galactic environments in a future paper.

\section*{Acknowledgements}
We thank the referee for a thorough report which improved this paper.
AAA, CLD, TJRB, and ASMB acknowledge funding from the European Research Council for the Horizon 2020 ERC consolidator grant project ICYBOB, grant number 818940. This work was performed using the DiRAC Data Intensive service at Leicester, operated by the University of Leicester IT Services, which forms part of the STFC DiRAC HPC Facility (www.dirac.ac.uk). The equipment was funded by BEIS capital funding via STFC capital grants ST/K000373/1 and ST/R002363/1 and STFC DiRAC Operations grant ST/R001014/1. DiRAC is part of the National e-Infrastructure. Figures were produced using \textsc{splash} \citep{price2007splash}, NumPy \citep{numpy}, Matplotlib \citep{matplotlib}, and Pandas \citep{pandas}.

\section*{Data availability}
The data underlying this paper will be shared on reasonable request to the corresponding author.



\bibliographystyle{mnras}
\bibliography{refs}



\appendix

\section{Effect of pixel size, accretion radius, and SFE on the Kennicutt-Schmidt relation}
\label{sec:pixelsize}
\begin{figure}
    \centering
	\includegraphics[width=0.95\columnwidth]{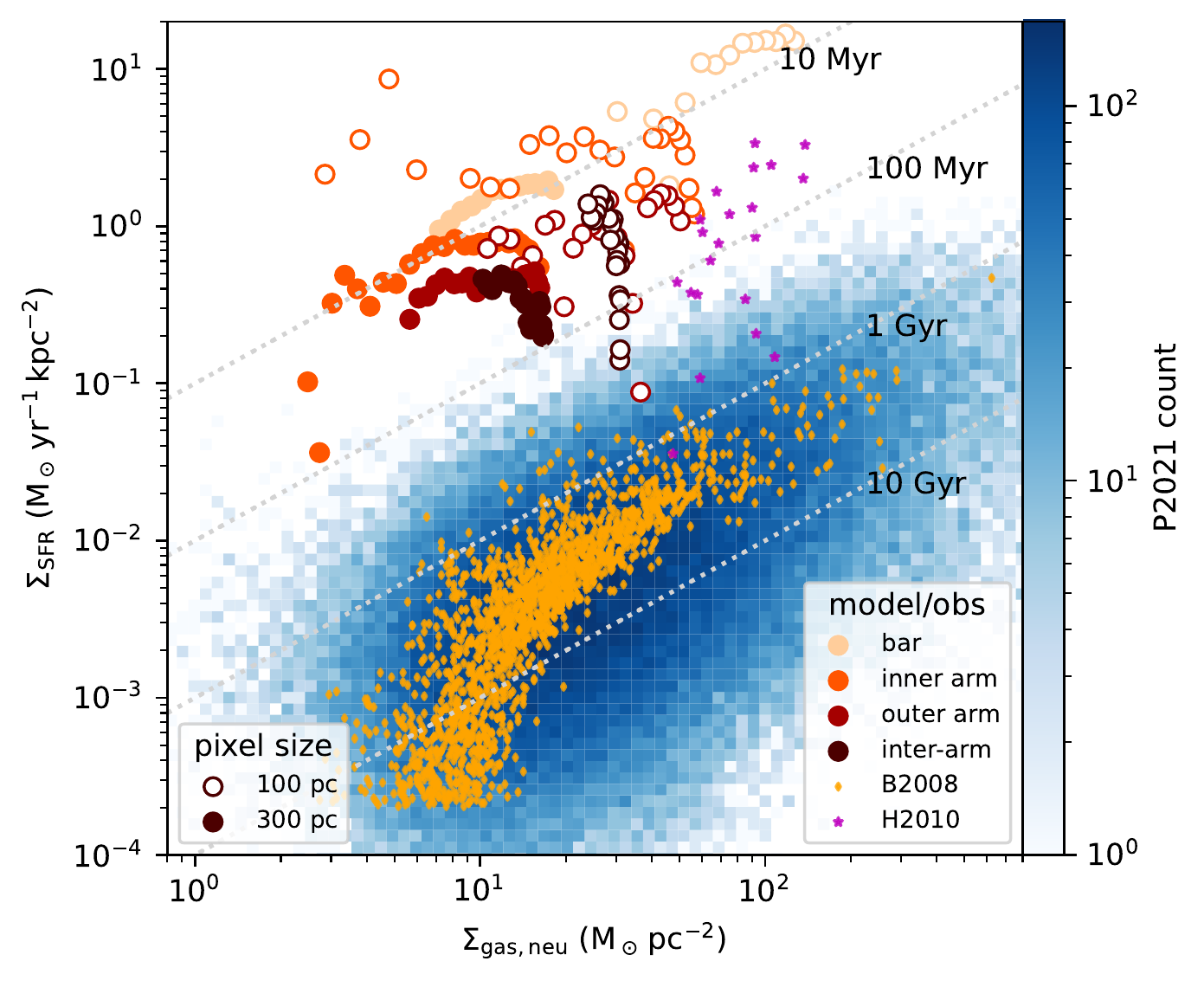}
    \caption{Same as the top panel of \cref{fig:kennicutt}, but now quantities are calculated inside squares of fixed size.}
    \label{fig:pixelsize}
\end{figure}
\begin{figure}
    \centering
	\includegraphics[width=0.95\columnwidth]{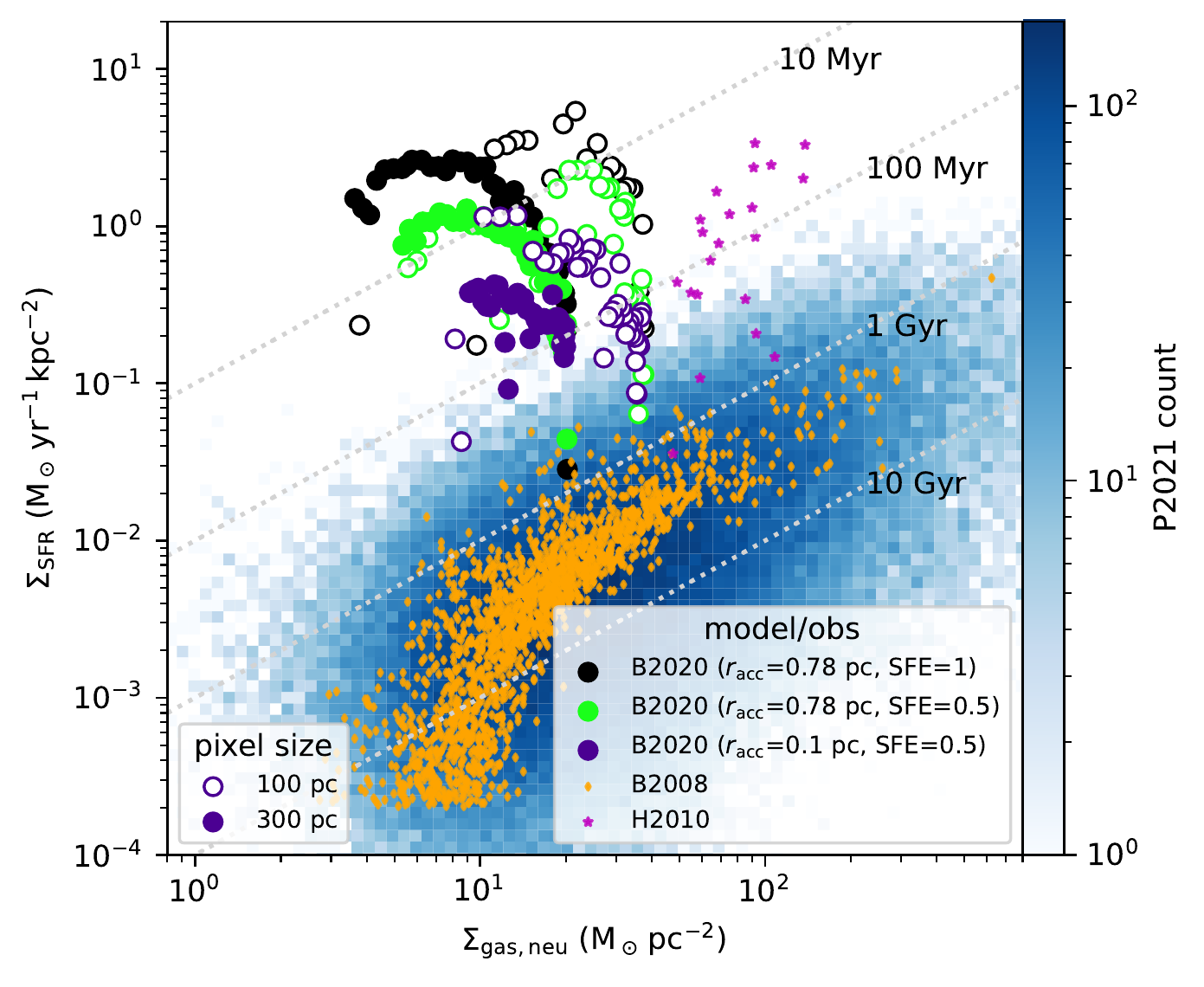}
    \caption{Same as \cref{fig:pixelsize} and the top panel of \cref{fig:kennicutt}, but now the models show the spiral arm region from \citet[][B2020]{bending2020} with different sink accretion radii ($r_\mathrm{acc}$) and different efficiencies for converting sinks to stars (SFE).}
    \label{fig:accretionradius}
\end{figure}

We recalculate $\sigmasfr$ and $\sigmagas$ in a fixed area with dimensions $X=Y= \SI{100}{\pc}$, instead of recalculating the actual area each time as we did for \cref{fig:kennicutt}. We repeat this for \SI{300}{\pc}. We calculate the mass of sinks and neutral gas in this area around the centre of mass, and repeat the procedure in \cref{sec:sfr} (with the exception that here we average the SFR over a longer $\Delta t \approx \SI{0.094}{\mega\yr}$, to reduce the number of points in the plot for clarity). The results are shown in \cref{fig:pixelsize} with open circles for the smaller area and filled circles for the larger area. Points for the smaller area are shifted towards the top right compared to the larger area, and are close to the Milky Way results of \citet{heiderman2010}. There is also more scatter, as sinks move in and out of the area between time steps, and this changes $\Delta M_\mathrm{sinks}$ in \cref{eq:sigmasfr}. For the 300\,pc area, the bar results now have systematically higher $\sigmasfr$ for the same $\sigmagas$ than the other regions (whereas for the 100\,pc area, and in \cref{fig:kennicutt}, they have higher $\sigmagas$ too) -- i.e. the bar points are vertically higher than the arms and inter-arm, not horizontally. This is because the bar is smaller than 300\,pc and hence is not resolved with this pixel size, whereas it is resolved with the 100\,pc pixel.

We also test the effect of sink accretion radius, $r_\mathrm{acc}$, in different runs of the spiral arm region from \citet{bending2020}. One run has $r_\mathrm{acc}$ = 0.78\,pc and one has $r_\mathrm{acc}$ = 0.1\,pc.
The results are shown in \cref{fig:accretionradius} with green circles showing the larger $r_\mathrm{acc}$ and purple circles the smaller $r_\mathrm{acc}$. Smaller $r_\mathrm{acc}$ results in lower SFR and brings the depletion time scales into agreement with the Milky Way regions. As before, the pixel size determines the diagonal position. 

Lastly, we test the star formation efficiency (SFE) imposed on sink particles, i.e. the proportion of each cluster-sink which is available for conversion to stars as described in \cref{sec:clustersinks}. The black circles in \cref{fig:accretionradius} have SFE=1 and can be compared with the green points where SFE=0.5. The latter points again have lower SFR.

Combined, these tests show that the K-S relation is sensitive to pixel size (diagonal position), SFE per sink (lower SFE gives lower $\sigmasfr$), and sink accretion radius (lower $r_\mathrm{acc}$ gives lower $\sigmasfr$).


\bsp	
\label{lastpage}
\end{document}